%% file: main_survey.tex
\documentclass[journal,comsoc]{IEEEtran}
\usepackage[T1]{fontenc}

\usepackage{cite}
\usepackage{color}
\usepackage{tabu}
\usepackage{subfigure}
\usepackage{amsmath}
\usepackage[cmintegrals]{newtxmath}
\usepackage{enumitem}

\ifCLASSOPTIONcompsoc
  \usepackage[caption=false,font=normalsize,labelfont=sf,textfont=sf]{subfig}
\else
  \usepackage[caption=false,font=footnotesize]{subfig}
\fi

\usepackage{graphicx}
\usepackage{caption}
\usepackage{url}
\usepackage{mathtools}
\usepackage{calc}
\usepackage{xparse}
\usepackage{array}
\usepackage{longtable}
\usepackage{multirow}
\usepackage{multicol,afterpage,wrapfig}
\usepackage[utf8]{inputenc}
\usepackage{adjustbox}
\usepackage{rotating}
\usepackage{soul}

\newcommand{\tabincell}[2]{\begin{tabular}{@{}#1@{}}#2\end{tabular}}

\newcommand{\bi}{\begin{itemize}}
	\newcommand{\ei}{\end{itemize}}
\newcommand {\beq}{\begin{equation}}
	\newcommand {\eeq}{\end{equation}}
\newcommand {\be}{\begin{enumerate}}
	\newcommand {\ee}{\end{enumerate}}

\newcounter{AFNumberOfComments}
\stepcounter{AFNumberOfComments}
\newcounter{MDNumberOfComments}
\stepcounter{MDNumberOfComments}
\newcounter{MHNumberOfComments}
\stepcounter{MHNumberOfComments}

\newcommand {\revise}{\color{black}}

\usepackage{pifont}
\newcommand{\cmark}{\ding{51}}%
\newcommand{\xmark}{\ding{55}}%
\input{macros}

\begin{document}

\title{Survey on UAV Cellular Communications: \\ Practical Aspects, Standardization Advancements, Regulation, and Security Challenges}

\author{Azade~Fotouhi,~\IEEEmembership{Student Member,~IEEE,}
		Haoran~Qiang,~\IEEEmembership{Student Member,~IEEE,} 
	Ming~Ding,~\IEEEmembership{Senior Member,~IEEE,}
	Mahbub~Hassan,~\IEEEmembership{Senior Member,~IEEE,}
	\IEEEauthorblockN{{Lorenzo~Galati~Giordano,~\IEEEmembership{Member,~IEEE,}  Adrian~Garcia-Rodriguez},~\IEEEmembership{Member,~IEEE,}}
	and~Jinhong~Yuan,~\IEEEmembership{Fellow,~IEEE} 
	\IEEEcompsocitemizethanks{\IEEEcompsocthanksitem A. Fotouhi is with School of Computer Science and Engineering, University of New South Wales (UNSW), Sydney, Australia, and also with Altran Technologies, France (E-mail: a.fotouhi@unsw.edu.au).
		\IEEEcompsocthanksitem H. Qiang is with School of Electrical Engineering and Telecommunications, University of New South Wales (UNSW), Sydney, Australia, and also with the Data61, CSIRO, Eveleigh, Australia (E-mail: h.qiang@student.unsw.edu.au).
		\IEEEcompsocthanksitem M. Hassan is with School of Computer Science and Engineering, University of New South Wales (UNSW), Sydney, Australia, and also with the Data61, CSIRO, Eveleigh, Australia (E-mail: mahbub.hassan@unsw.edu.au).
		\IEEEcompsocthanksitem M. Ding is with the Data61, CSIRO, Eveleigh, Australia (E-mail: ming.ding@data61.csiro.au).
		\IEEEcompsocthanksitem L.~Galati~Giordano, and A.~Garcia-Rodriguez are with Nokia Bell Labs, Ireland (E-mail: lorenzo.galati\_giordano, adrian.garcia\_rodriguez@nokia-bell-labs.com).
		\IEEEcompsocthanksitem J. Yuan is with School of Electrical Engineering and Telecommunications, University of New South Wales (UNSW), Sydney, Australia (E-mail: j.yuan@unsw.edu.au).}
}


\markboth{Journal of Communications Surveys and Tutorials}%
{UAVs in Cellular Networks}

\maketitle

\begin{abstract}
The rapid growth of consumer Unmanned Aerial Vehicles (UAVs) is creating promising new business opportunities for cellular operators. 
On the one hand, 
UAVs can be connected to cellular networks as new types of user equipment, 
therefore generating significant revenues for the operators that can guarantee their stringent service requirements. 
On the other hand, 
UAVs offer the unprecedented opportunity to realize UAV-mounted flying base stations that can dynamically reposition themselves to boost coverage, 
spectral efficiency, 
and user quality of experience. 
Indeed, 
the standardization bodies are currently exploring possibilities for serving commercial UAVs with cellular networks. 
Industries are beginning to trial early prototypes of flying base stations or user equipments, 
while academia is in full swing researching mathematical and algorithmic solutions to address interesting new problems arising from flying nodes in cellular networks. 
In this article, 
we provide a comprehensive survey of all of these developments promoting smooth integration of UAVs into cellular networks. 
Specifically, 
we survey (i) the types of consumer UAVs currently available off-the-shelf, 
(ii) the interference issues and potential solutions addressed by standardization bodies for serving aerial users with the existing terrestrial base stations, 
(iii) the challenges and opportunities for assisting cellular communications with UAV-based flying relays and base stations, 
(iv) the ongoing prototyping and test bed activities, 
(v) the new regulations being developed to manage the commercial use of UAVs, 
and (vi) the cyber-physical security of UAV-assisted cellular communications.
\end{abstract}


\IEEEpeerreviewmaketitle

\input{Introduction}

\input{UAVTypes}

\input{Standardization}
\input{AerialBaseStations}
\input{Prototyping}

\input{Regulation}

\input{Security}
\input{LessonLearned}

\input{FutureDirection}

\input{Conclusion}

\ifCLASSOPTIONcaptionsoff
  \newpage
\fi
%
\bibliographystyle{unsrt}
\bibliography{surveybib}

\end{document}

%% file: macros.tex
\newfont{\bbb}{msbm10 scaled 500}

\newfont{\bb}{msbm10 scaled 1100}

\newcommand{\executeiffilenewer}[3]{%
\ifnum\pdfstrcmp{\pdffilemoddate{#1}}%
{\pdffilemoddate{#2}}>0%
{\immediate\write18{#3}}\fi%
}

%% file: Introduction.tex
\section{Introduction}
\label{sec:litreview_Introduction}

From aerial photography to search-and-rescue to package delivery --- the use cases of consumer unmanned aerial vehicles (UAVs) (a.k.a. drones) are exploding. 
According to a report from Federal Aviation Administration (FAA), the fleet of drones will be more than doubled from an estimated 1.1 million vehicles in 2017 to 2.4 million units by 2022~\cite{statistics4}.
It is expected that new use cases will continue to emerge, 
fuelling further growth in UAVs. 
As many of these use cases would benefit from connecting the UAVs to the cellular networks for better control and communications, 
the growth in the UAV market is expected to bring new promising business opportunities for cellular operators.    

The Third Generation Partnership Project (3GPP), 
which oversees the standards activities for cellular networks, 
has recently concluded a study item~\cite{3GPP36777} to explore the challenges and opportunities for serving the UAVs as a new type of user equipment (UE), 
referred to as~\textit{aerial UE}. 
An interesting finding of this study is that the enhanced line-of-sight (LOS) between aerial UE and ground base stations (BSs) would significantly increase interference in the system, 
which calls for new strategies to seamlessly accommodate both aerial and ground UEs in the same system. 
A variety of new techniques have already been proposed to address such interference issue, 
which show promising results. 

While 3GPP is mainly concerned with connecting UAVs to cellular networks, 
industry and academia are advancing to the next level of research and development that promises to harness the full potential of UAVs communications. 
In particular, 
they are exploring the unprecedented opportunity to realize UAV-mounted flying relays and BSs that can dynamically reposition themselves to boost coverage, 
spectral efficiency, 
and user quality of experience (QoE). 
Major vendors have already field-trialled their prototypes to demonstrate the proof-of-concept of such UAV-mounted flying BSs~\cite{nokiadrone, TaxiDrone}. 
A large number of papers have been published in recent years proposing novel algorithms to optimize positioning and mobility of flying relays and BSs \cite{our_ieeeaccess,8422706,2018arXiv180300680M}.



Given the significant momentum and recent activities promoting UAV in cellular networks, 
it is timely to survey this brand-new field. 
Although several survey articles on UAV have been published in recent years, 
none of them had focused on \textit{the practical aspects of cellular UAV communications}. 
For example, 
the surveys in~\cite{7317490,7208426} focussed on forming ad hoc networks between many UAVs in the sky. 
Hayat et al.,~\cite{7463007} surveyed communications demands for various applications of UAV and analysed the suitability of existing wireless technologies, 
including Bluetooth, Zigbee, Wi-Fi, WiMAX, and cellular, to meet these demands. 
In a magazine paper, 
Sekander et al.,~\cite{ikram2017tier} analyzed the opportunities for drones to assist cellular networks, 
but specialized on combining drones from different altitudes to form a multi-tier drone network. 
In another magazine paper, 
Zeng et al.,~\cite{zeng2016wireless} surveyed the issues and opportunities for using drones to assist wireless networks in general without specific focus on cellular networks. 
Recently, 
Mozaffari et al.,~\cite{2018arXiv180300680M} delivered a comprehensive tutorial on UAV wireless communications. 
Our work complements their vision by covering a variety of cellular-specific issues such as the relevant 3GPP developments, 
vendor prototypes of flying BSs, 
regulations and cyber-security issues affecting cellular UAVs, 
and the potential impacts of UAV adoption on the cost and business models of cellular networks.  

\begin{figure}[t!]
	\centering
	\includegraphics[width=1\linewidth]{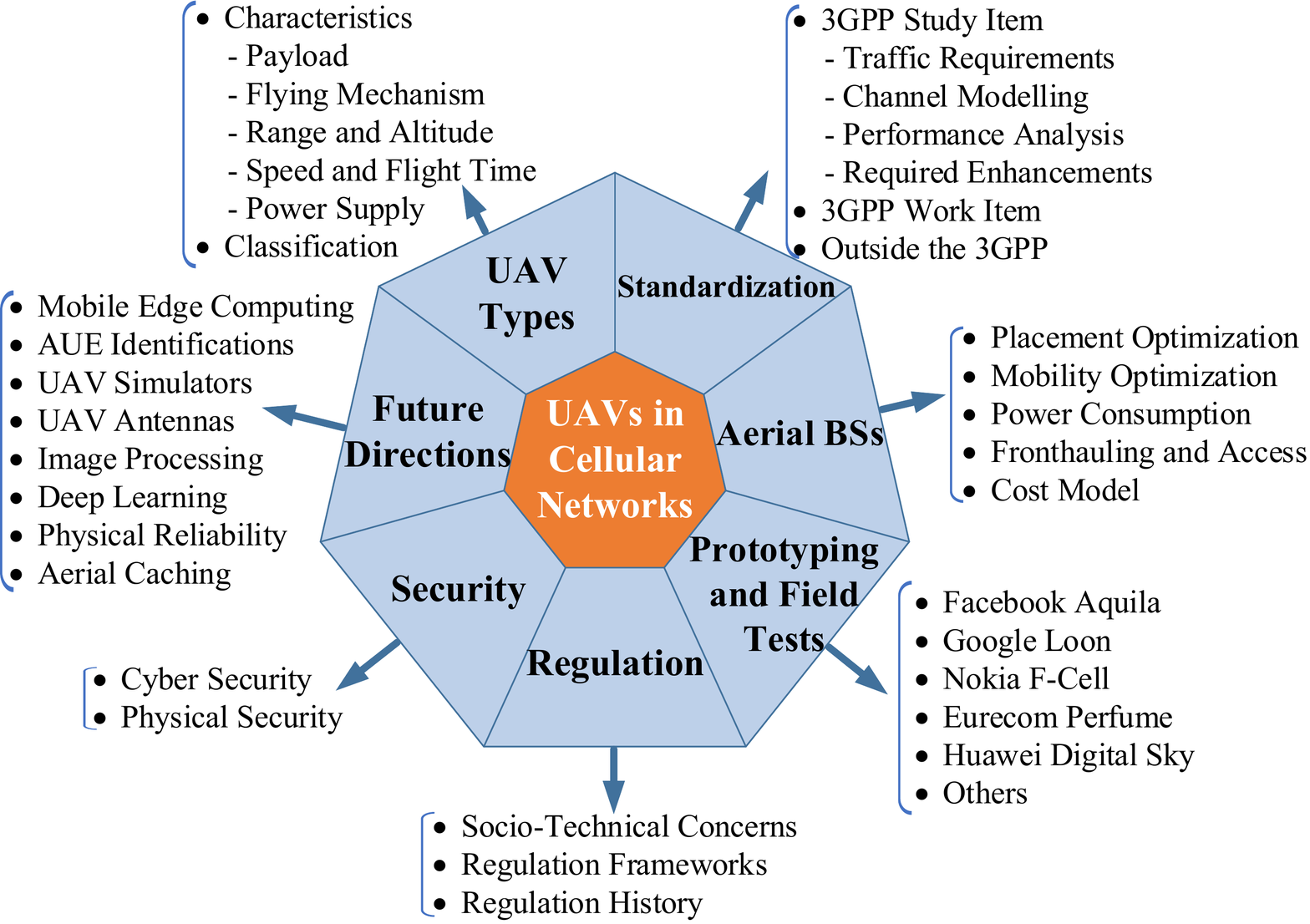}
	\caption{Taxonomy of the survey}
	\label{fig:taxonomy}
\end{figure}

\begin{figure*}[!t]
	\center   
	\subfigure[Parrot Swing (front view).]{\includegraphics[width=0.3\linewidth]{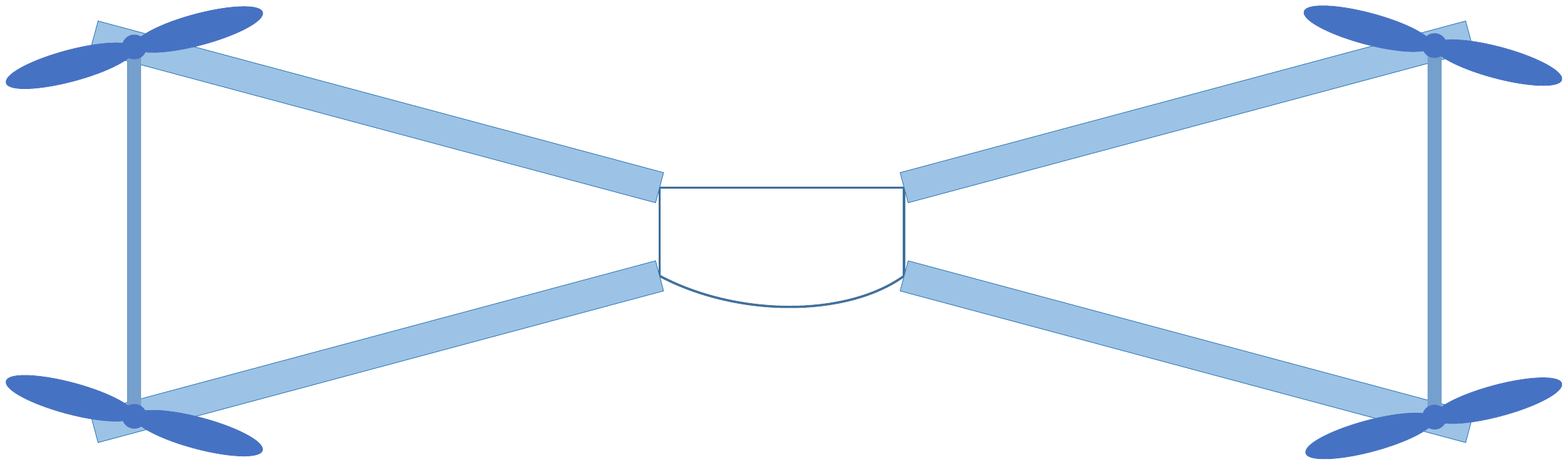}}
	\subfigure[Kogan Nano Drone.]{\includegraphics[width=0.3\linewidth]{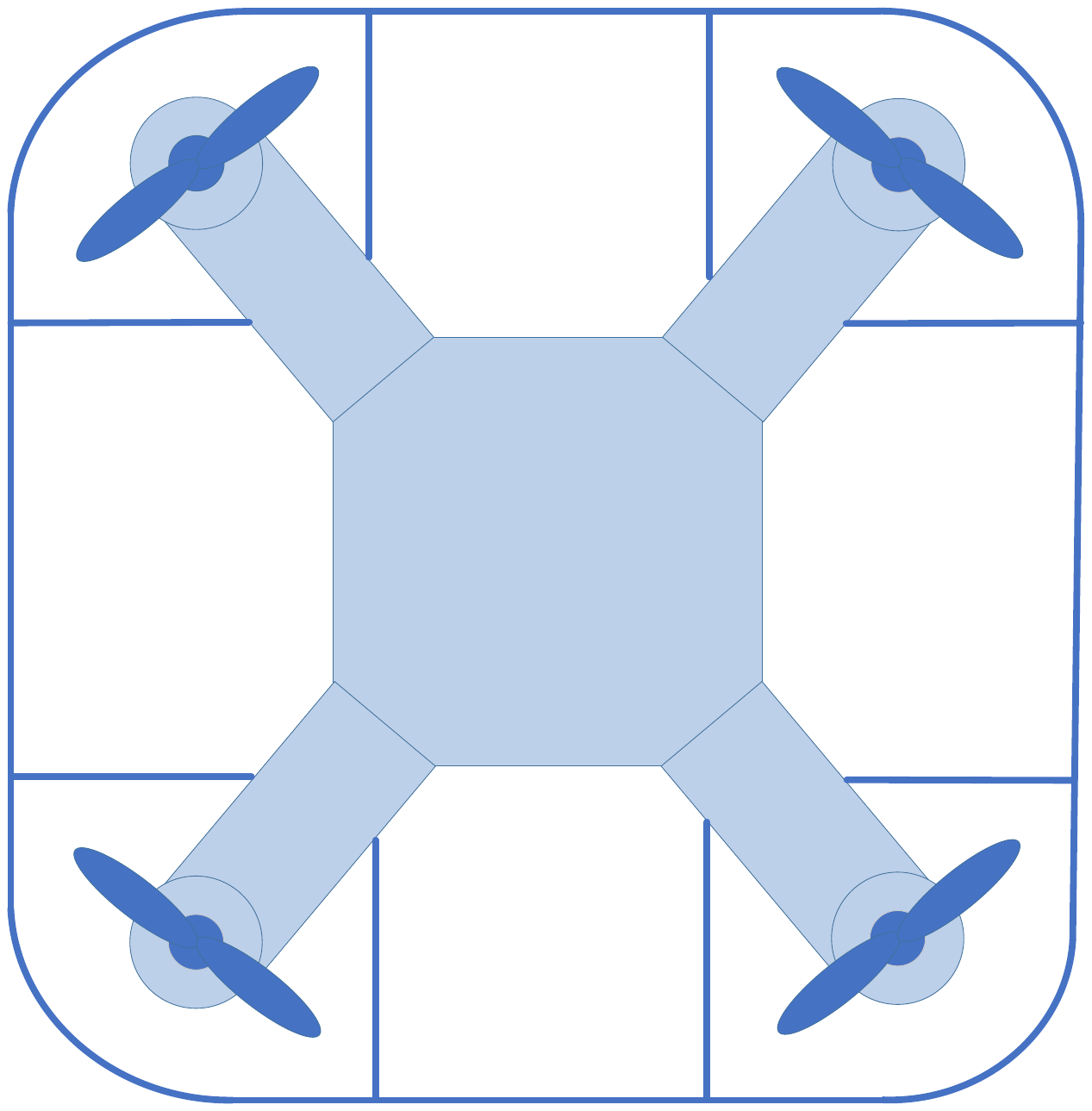}}
	\subfigure[Parrot Disco.]{\includegraphics[width=0.3\linewidth]{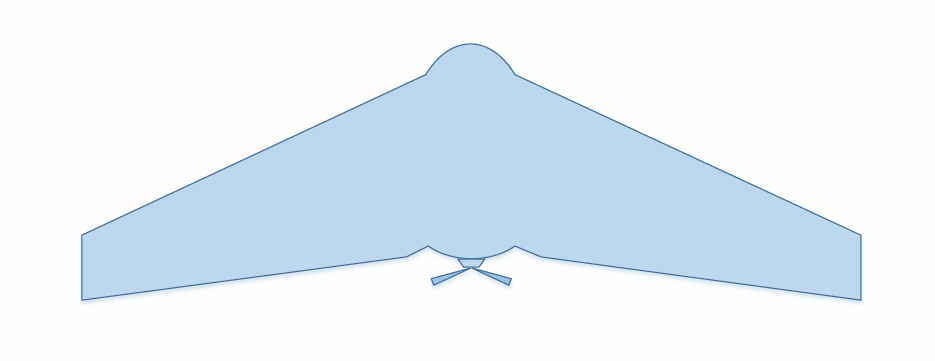}}
	\subfigure[DJI Spreading Wings S900.]{\includegraphics[width=0.3\linewidth]{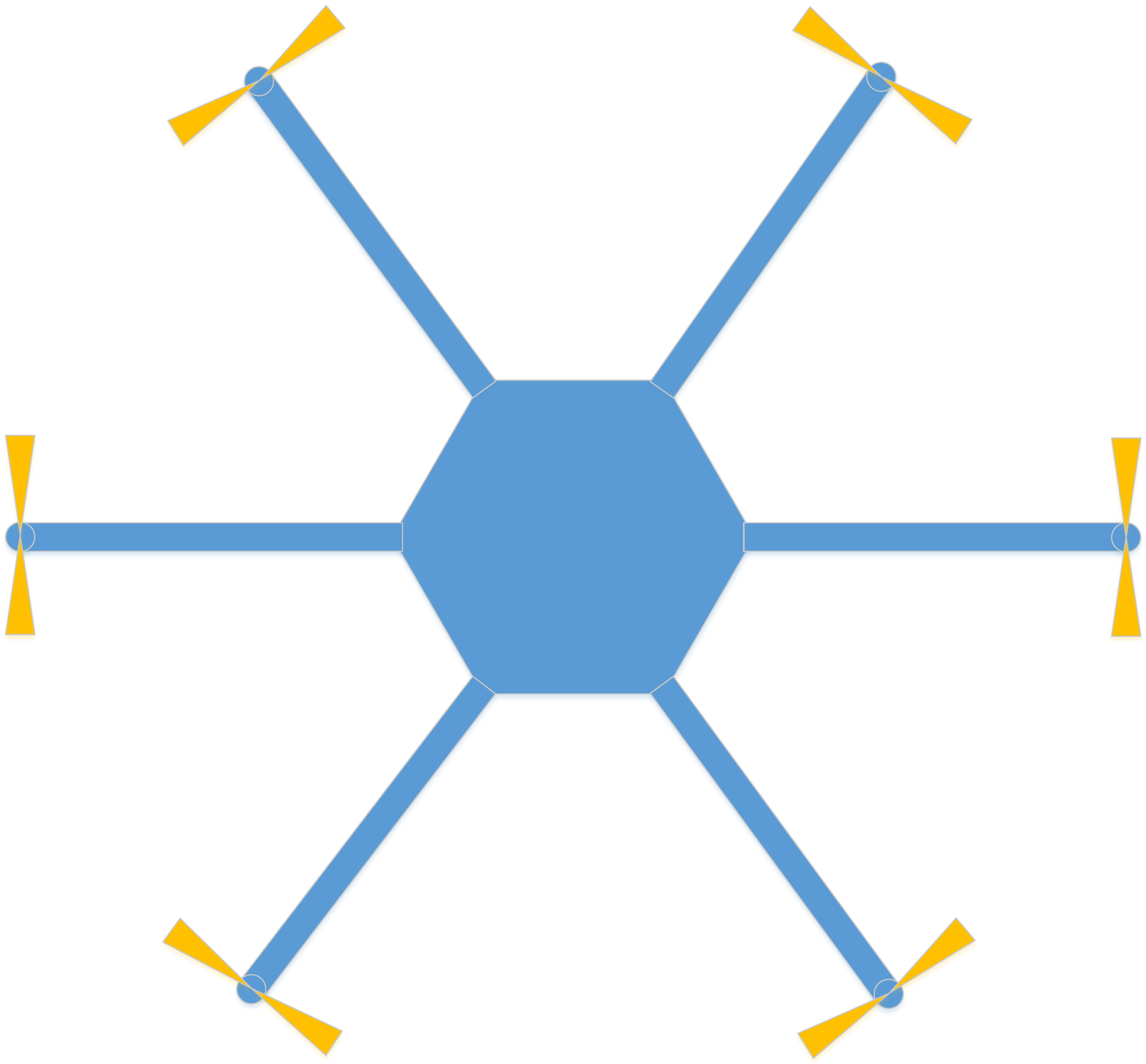}}
     \subfigure[Scout B-330 UAV helicopter.]{\includegraphics[width=0.3\linewidth]{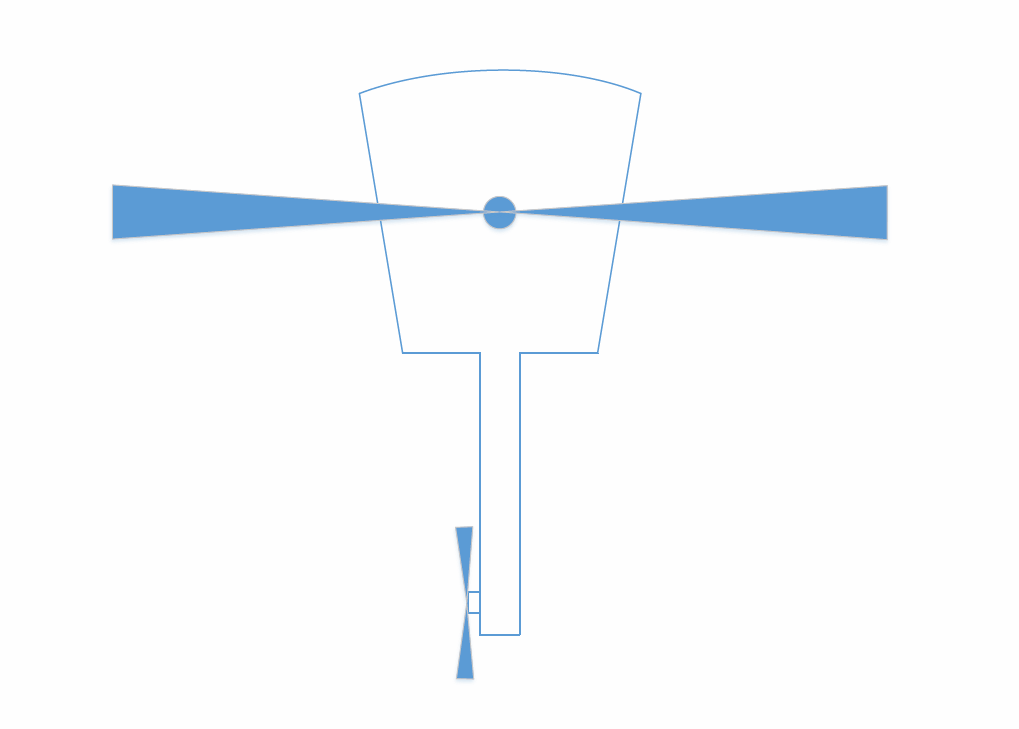}}
	\subfigure[Predator B.]{\includegraphics[width=0.3\linewidth]{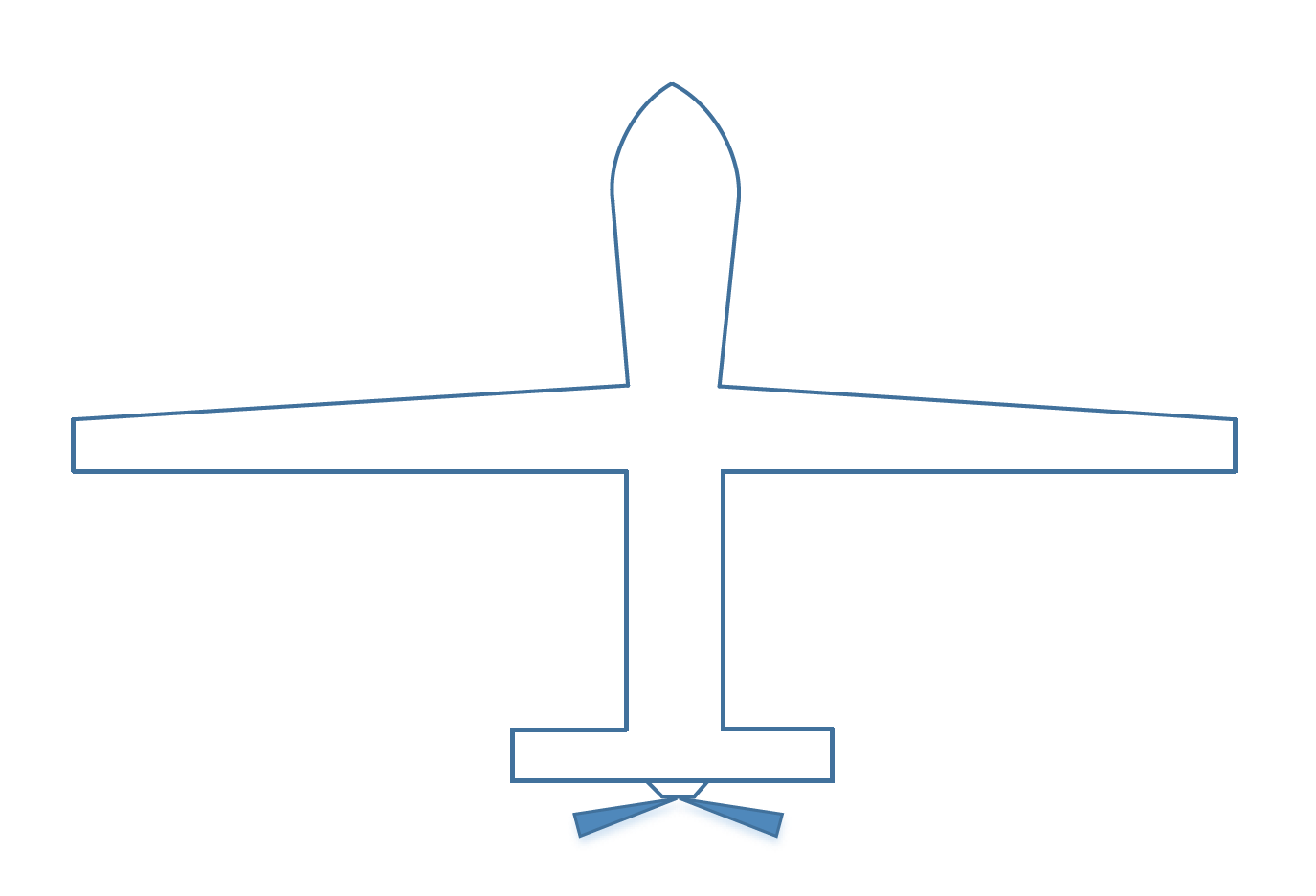}}
	\caption{Images of different UAV types.}
	\label{fig:uavtype}
	\vspace{-0.3cm}
\end{figure*}

A graphical illustration of the detailed taxonomy of our survey is presented in Fig. \ref{fig:taxonomy}. 
{\revise{This paper consists of seven main sections where the sections' headings are displayed in the second level in Fig.~\ref{fig:taxonomy}. 
Accordingly, 
the third level represents the subsections under each section. }}
{\revise{Specifically, 
we survey the types of consumer UAVs currently available off-the-shelf highlighting their potential roles within cellular networks (Section \ref{s:uavs}), 
the 3GPP developments regarding interference issues and solutions for serving aerial users (Section \ref{s:standardization}), 
the challenges and opportunities for assisting cellular communications with UAV-mounted flying base stations or relays (Section \ref{s:flying}), 
the most advanced UAV prototyping and field trial examples (Section \ref{s:DronePrototypingAndFieldTests}), 
the new regulations being developed to manage the commercial use of UAVs (Section \ref{s:reg}), 
the cyber-physical security issues for UAV-assisted cellular communications (Section \ref{s:security}), 
and finally an overlook of the most promising future research directions (Section \ref{s:future}) followed by the conclusions (Section \ref{sec:conclusion}).}}

%% file: UAVTypes.tex
\begin{table*}[!t]
	\centering
	\caption{Characteristics of different drone types}
	\label{table:drone_type}
	\begin{tabular}{|m{1.5cm}<{\centering}|m{2.8cm}<{\centering}|m{2.8cm}<{\centering}|m{2.8cm}<{\centering}|m{2.8cm}<{\centering}|m{2.8cm}<{\centering}|}
		\hline
		& \textbf{Micro (weight$\leqslant$100g)} & \textbf{Very Small (100g$<$weight$<$2kg)} & \textbf{Small (2kg$\leqslant$weight$<$25kg)} & \textbf{Medium (25kg$\leqslant$weight$\leqslant$150kg)} & \textbf{Large (weight$>$150kg)} \\ \hline
		\textbf{Model} & Kogan Nano Drone & Parrot Disco & DJI Spreading Wings S900 & Scout B-330 UAV helicopter & Predator B\\ \hline
		\textbf{Ref.} & \cite{Kogannano} & \cite{Parrotdisco} & \cite{DJIS900} & \cite{Scout} & \cite{predatorB} \\ \hline
		\textbf{Illustration} & Fig. \ref{fig:uavtype}(b) & Fig. \ref{fig:uavtype}(c) & Fig. \ref{fig:uavtype}(d) & Fig. \ref{fig:uavtype}(e) & Fig. \ref{fig:uavtype}(f) \\ \hline
		\textbf{Weight} & 16g & 750g & 3.3kg & 90kg & 2223kg \\ \hline
		\textbf{Payload} & N/A & N/A & 4.9kg & 50kg & 1700kg \\ \hline
		\textbf{Flying Mechanism} & Multi-rotor & Fixed-wing & Multi-rotor & Multi-rotor & Fixed-wing \\ \hline
		\textbf{Range} & 50-80m & 2km & N/A & N/A & 1852km \\ \hline
		\textbf{Altitude} & N/A & N/A & N/A & 3km & 15km \\\hline
		\textbf{Flight Time}  & 6-8min & 45min & 18 min & 180min & 1800min \\ \hline
		\textbf{Speed} & N/A & 80km/h & 57.6km/h & 100km/h (horizontal) & 482km/h \\ \hline
		\textbf{Power Supply} & {\revise{3.7V/}}160mAh Li-battery & 2700mAh{\revise{/25A}} 3-cell LiPo Battery & LiPo Battery {\revise{(6S, 10000mAh$\sim$15000mAh, 15C(Min))}} & {\revise{Gasoline (heavy fuel optional)}} & {\revise{950-shaft-horsepower Turboprop Engine}} \\ \hline
		\textbf{\revise{Power Consumption}} & {\revise{N/A}} & {\revise{N/A}} & {\revise{Maximum: 3kW; Hover: 1kW}} & {\revise{Engine: 21kW; Onboard power generator for payload: 1.5kW}} & {\revise{Engine: 712kW}} \\ \hline
   	    \textbf{Application} & Recreation; {\revise{suitable to carry sensors for indoor wireless data collection}} & Recreation; suitable to carry cellular UEs &  Professional aerial photography and cinematography; suitable to carry cellular BSs or UEs & Survey (data acquisition), HD video live stream; {\revise{suitable to carry or act as motorial energy source for wireless energy transfer, and act as aerial caches;}} can carry cellular BSs or UEs  & Armed reconnaissance, airborne surveillance, and target acquisition \\ \hline
	\end{tabular}
\end{table*}

\section{\revise{UAVs Types and Characteristics}}
\label{s:uavs}

UAVs, 
commonly known as drones, 
are available in different sizes and specifications, 
as illustrated in Fig.~\ref{fig:uavtype}. 
They can be deployed quickly whenever needed, 
which makes them promising candidates for providing cellular connectivity. 
In this section, 
the characteristics and features of a few typical drones are summarized and explained (see Table \ref{table:drone_type}), 
with special focus on their impact on UAV-aided cellular communications.





\subsection{Payload}

Payload refers to the maximum weight that a drone can carry, 
which measures its lifting capability. 
Payloads of drones vary from tens of grams up to hundreds of kilograms~\cite{tang2015drone}. 
The larger the payload, 
the more equipment and accessories can be carried at the expense of a larger drone size, 
higher battery capacity, 
and shorter duration in the air. 
Typical payloads include video cameras and all sorts of sensors, 
which could be used for reconnaissance, 
surveillance and commercial purposes~\cite{fahlstrom2012introduction}. 
When assisting cellular communications, 
drones can carry cellular UEs such as mobile phones or tablets, 
whose weight is usually less than 1 kilogram~\cite{8048502}. 
BSs or remote radio heads (RRHs) can also be carried by or mounted on drones to provide cellular services. 
In this case, 
payload of drones should be at least a few kilograms. 

\subsection{Flying Mechanism} 

Depending on their flying mechanisms, 
drones can be classified into three types: 

\begin{itemize}
\item \emph{Multi-rotor drones} (also known as rotary-wings drones) allow vertical take-off and landing, 
and can hover over a fixed location to provide continuous cellular coverage for certain areas. 
This high manoeuvrability makes them suitable for assisting cellular communications, 
since they can deploy BSs at the desired locations with high precision, 
or fly in a designated trajectory while carrying BSs. 
However, 
multi-rotor drones have limited mobility and consume significant power as they have to fight against gravity all the time.
\item \emph{Fixed-wing drones} can glide over the air, 
which makes them significantly more energy efficient and able to carry heavy payload. 
Gliding also helps fixed-wing drones to travel at a faster speed. 
The downsides of fixed-wing drones are that (i) they require a runway to take off and land as vertical take-off and landing are not possible~\cite{nonami2010introduction}, 
and (ii) they cannot hover over a fixed location. 
Fixed-wing drones are also more expensive than multi-rotor drones.
\item \emph{Hybrid fixed/rotary wing drones} have recently reached the market to provide a compromise between the two above-mentioned drone types. 
An illustrative example of a hybrid fixed/rotary wing drone is the Parrot Swing shown in Fig.~\ref{fig:uavtype}(a), 
which can take off vertically, 
quickly reach its destination by gliding through the air, 
and then switch to hovering using four rotors. 
\end{itemize}


\subsection{Range and Altitude}

The range (one hop) of a drone refers to the distance from which it can be remotely controlled. 
The range varies from tens of meters for small drones to hundreds of kilometers for large ones. 
Altitude here refers to the maximum height a drone can reach regardless of the country-specific regulations. 
The maximum flying altitude of a given drone is a critical parameter for UAV-aided cellular communications, 
since a UAV BS needs to vary its altitude to maximize the ground coverage and satisfy different quality of service (QoS) requirements~\cite{6863654}. 
Overall, 
aerial platforms can be classified into two types depending on their altitude:
\begin{itemize}
\item \emph{Low-altitude platforms (LAPs)} are usually employed to assist cellular communications since they are more cost-effective and allow fast deployment. 
Moreover, 
LAPs usually provide short-range line-of-sight (LOS) links that can significantly enhance the communication performance~\cite{zeng2016wireless, 7335646, Ding2017ULlos}.
\item \emph{High-altitude platforms (HAPs)} such as balloons can also provide cellular connectivity~\cite{zeng2016wireless}. 
Compared to LAPs, 
HAPs have a wider coverage and can stay much longer in the air. 
{\revise{However, 
HAP deployment is more complex and they are mainly considered as a vehicle to provide Internet connectivity to large fractions of world population currently not served by cellular networks.
More importantly, 
using HAPs in cellular communications may cause total network outage due to extremely large inter-cell interference~\cite{7842150, Ding2016ASECrash}.   
As such, 
they are rarely considered in the literature on UAV-aided cellular networks, 
but rather pursued by Internet companies. 
In Section~\ref{s:DronePrototypingAndFieldTests}, 
we discuss examples of HAP introduced by Google and Facebook.}}
\end{itemize}

\subsection{Speed and Flight Time} 

Small drones typically travel at speeds below 15\,m/s~\cite{quintero2010optimal}, 
while large drones can reach an impressive speed of 100\,m/s~\cite{zeng2016wireless}. 
When a UAV BS/relay flies in a designated trajectory to maximize its energy and spectral efficiency, 
its speed needs to be carefully considered if the trajectory requires frequent turns.
The trade-off between a drone's speed and its turning agility is studied in~\cite{7974336}. 
The maximum time a drone can spend in the air without recharging or refueling is referred to as its flight time or endurance. 
Small commercial drones usually have a flight time of 20-30 minutes, 
while some large drones can last for hours~\cite{sayler2015world}. 
Emerging technologies have prolonged the endurance of small drones. 
For example, 
the Skyfront Tailwind drone can achieve an endurance of up to 4.5 hours with hybrid-electric power sources~\cite{Skyfront}. 
Nevertheless, 
the limited endurance of the existing off-the-shelf UAVs is currently one of the major practical factors restricting their full-scale deployment in cellular networks.

\subsection{Power Supply}

A drone's power supply significantly determines its endurance. 
While rechargeable batteries power most commercial drones, 
some large drones can be powered by fuels such as gas for longer flight times~\cite{dudek2013hybrid}. 
Employing solar energy to power drones is also a promising technique~\cite{shaheed2015flying}. 
For drone-mounted BSs, 
power supply needs to support the functionality of both the drone and its on-board equipment such as antenna array, amplifier, circuits, etc. 
{\revise{For example, a typical aerial BS requires 5\,W as its maximum transmit power \cite{sekander2018multi, wang2018power, Ding2016TWC_IMC}, 
which should be supplied by its on-board energy source.}}    

\subsection{An Example of Drone Classification}

Civil aviation authorities usually classify drones based on their gross weights. 
Table~\ref{table:drone_type} demonstrates such weight-based classification as adopted by Civil Aviation Safety Authority (CASA) Australia \cite{casaadvisory} by listing typical drones and their features, 
which are depicted in Fig. \ref{fig:uavtype}(b)$-$(f).

%% file: Standardization.tex
\section{ \revise{Standardization: \\ Enabling UAV Cellular Communications}} \label{s:standardization}
{\revise{{The cellular industry has recognized the importance of providing support to low-altitude UAVs for enabling beyond LOS control and establishing a reliable communication~\cite{3GPP36777, lin2017sky, BerChiPol2016, GerGarUAVICC2018, GarGerGal_WCM_2018}. 
This section summarizes the outcomes and current status of a number of industry-led initiatives with the above targets.}}}

\subsection{\revise{3GPP Study Item Phase}}
\label{subsec:3GPPSI}

The third generation partnership group (3GPP) defined a study item (SI) in March 2017 with four fundamental objectives: 
i) the understanding of the UAV traffic requirements; 
ii) the development of a channel model to characterize air-to-ground propagation characteristics; 
iii) the determination of whether the current LTE infrastructure could be reused to provide cellular service to aerial devices; 
iv) the definition of the enhancements required to effectively serve UAVs building up on LTE Release 14 functionalities~\cite{RP170779}. 
The SI finalized in December 2017 and the main results in each of the above four areas are summarized in the following~\cite{3GPP36777}.

\vspace*{0.2cm}
{\revise{\emph{i) 3GPP Study Item: UAV Traffic Requirements}}}
\vspace*{0.2cm}

The 3GPP identified the traffic types that cellular networks should cater for UAVs flying between ground level and 300 meters. 
These are summarized in Table~\ref{table:trafficRequirements3GPP} and can be classified into three categories: 
1) synchronization and radio control, 2) command \& control, and 3) application data. 
\begin{enumerate}
\item The information contained within the synchronization and radio control messages is essential for a successful association and connectivity to the network. 
The transmission of these signals must be robust enough to guarantee that they can be decoded by flying UAVs. 
Examples of synchronization and radio control signalling include primary and secondary synchronization signals (PSS/SSS) and the physical downlink control channel (PDCCH), respectively.
\item Command \& control (C\&C) traffic enables beyond line-of-sight UAV piloting and has strict quality of service requirements (QoS) in terms of latency and reliability. 
Cellular operators have identified an attractive business opportunity in the management of this traffic, since it can be offered as a complementary network service to organizations interested in reliably controlling their UAVs.
\item While downlink data traffic is predominant in existing cellular communications, 
UAV application data transmissions are expected to be uplink-dominated. 
Transferral of live video streaming data and photos captured by camera-equipped UAVs contribute towards this traffic imbalance.
\end{enumerate}

\begin{table}[t]
\centering
\caption{UAV communication requirements~\cite{3GPP36777, GerGarUAVICC2018, 8647634}}
\label{table:trafficRequirements3GPP}
\renewcommand{\arraystretch}{1.5}
\begin{tabular}{|c|c|c|c|}\hline
\textbf{} & \multicolumn{1}{c|}{\textbf{Data Type}} & \multicolumn{1}{c|}{\textbf{Data Rate}} & \multicolumn{1}{c|}{\textbf{Critical?}} \\\hline
\multirow{3}{*}{DL} & {Synchronization (PSS/SSS)} & \multirow{2}{*}{N/A} & \cmark \\\cline{2-2}\cline{4-4}
& {Radio control (PDCCH)} &  & \cmark \\\cline{2-4}
 & \multicolumn{1}{c|}{{Command and control (C\&C)}} & \multicolumn{1}{c|}{60-100~kbps} & \cmark \\\cline{1-4}
\multirow{2}{*}{UL} & \multicolumn{1}{c|}{Command and control (C\&C)} & \multicolumn{1}{c|}{60-100~kbps} & \cmark \\\cline{2-4}
 & \multicolumn{1}{c|}{Application data} & \multicolumn{1}{c|}{Up to 50 Mbps} & \xmark \\\hline
\end{tabular}
\end{table}

\vspace*{0.2cm}
{\revise{\emph{ii) 3GPP Study Item: Channel Modelling}}}
\vspace*{0.2cm}


In order to characterize the performance of existing cellular networks when serving both ground and aerial devices, 
the 3GPP developed a UAV-specific statistical channel model building upon that defined in~\cite{3GPP38901}. 
This channel model complements those developed by the academic community. These academic models are summarized in Table~\ref{tbl:channel_exp} and Table~\ref{tbl:channel_simul} into three different categories ({\revise{Air to Ground (A2G)}}, Air to Air (A2A), and Ground to Air (G2A)) for completeness. 
The 3GPP-proposed models for rural-macro (RMa), urban-macro (UMa), and urban-micro (UMi) BS deployments are the result of a large number of measurement campaigns carried out by the standard-contributing companies and its main UAV-related features can be summarized as follows:
\begin{itemize}
\item \emph{UAV spatial placement:} The 3GPP defines five different cases depending on the density of UAVs in the network, 
i.e., it considers the deployment of $N_{\mathrm{aerial}} = \lbrace 0, 0.1, 1, 3, 5 \rbrace$ UAVs per cellular sector (out of a total of 15 mobile devices). 
Airborne devices are uniformly distributed between 0 and 300 meters and travel at a speed of 160 km/h.
\item \emph{LOS probability:} The LOS probability between ground BSs and UAVs grows as the latter increase their height. 
Remarkably, UAVs flying higher than 100 meters are considered to be in LOS with all the cellular BSs deployed in the network in the UMa scenario.
\item \emph{Path loss:} The 3GPP model also captures the fact that the path loss exponent of ground-to-aerial links generally decreases as UAVs increase their height. 
Indeed, UAVs in LOS with their BSs experience a path loss similar to of free-space propagation ($\alpha = 2.2$).
\item \emph{Shadowing:} The standard deviation of the log-normally distributed shadowing gain diminishes for increasing UAV heights, 
provided that the considered UAV-BS pair is LOS.
\item \emph{Fast-fading model:} Three different alternatives with a varying degree of implementation complexity are considered in~\cite{3GPP36777}, 
namely, 1) a variation of the cluster delay-based channel model developed in~\cite{3GPP38901} with UAV-specific channel characteristics such as the existence of a specular reflection on the building roof for the UMa scenario, 
2) an approach where the mean and standard deviation of the large scale parameters defined in~\cite{3GPP38901}  (delay spread, angular spreads of departure and arrival, and K-factor) are adjusted, 
and 3) a simpler alternative where, when compared with the channel model of~\cite{3GPP38901}, only the K-factor is adjusted.
\end{itemize}

\begin{table*}[]
	\centering
	\caption{Experimental studies for channel modeling}
	\label{tbl:channel_exp}
	\begin{tabular}{|m{0.7cm}|m{0.7cm}|m{2.5cm}|m{1.5cm}|m{1.5cm}|m{3cm}|m{3.5cm}|}
		\hline
		\textbf{Type}                 & \textbf{Ref.} & \textbf{Frequency/Protocol} & \textbf{Altitude} & \textbf{Environment} & \textbf{Experiment Details} & \textbf{Objective} \\ \hline
		\multirow{3}{*}{\textbf{A2G}} &    ~\cite{qualcommuavtr}          & PCS, AWS, and 700MHz     & below 120m    &   mixed suburban (California)  &  Custom designed quadrotor drone (5m/s)               & Enhance the understanding of aerial communications   \\ \cline{2-7} 
		&~\cite{7357682}        &   970Mhz (L-band), and 5060MHz (C-band)   &   560m  &   Near-urban (Cleveland) &          S-3B Viking aircraft         &    Characterizing the channel for air to ground communication           \\ \cline{2-7} 
		&~\cite{8048502}        &   {\revise{850Mhz, LTE }}&    {\revise{15,30,60,90, and 120 m }} &  {\revise{suburb in
		Victoria, Australia}} &       {\revise{ one commercial UAV, 4.8m/s, sony Xperia phone for logging   }}     &  {\revise{Modeling the excessive path loss exponent considering BS's down-tilted antennas   }}   \\ \cline{2-7} 
		&      ~\cite{7936620}        &    800MHz, LTE &    15,30,60, and 120 m &  Denmark   &    One commercial UAV , 15km/h            & Modeling the path loss exponent and shadowing       \\ \hline
		\multirow{3}{*}{\textbf{A2A}} &~\cite{7414180} &  2.4GHz, IEEE 802.11   &  Below 50m   &  -   & Two AscTec Firefly Hexacopter UAVs  &   Studying the impact of distance       \\ \cline{2-7} 
		&~\cite{7470935}                   &  ZigBee 802.145.4   &  Below 20m   &     -             &      Two Hexacopters   &   Measuring path loss exponent    \\ \hline 
		\multirow{2}{*}{\textbf{G2A}} &  ~\cite{5700251}           &  802.11b/g    &   75m  &     &  Fix-wing fuselag                &       Measuring the diversity in G2A links        \\ \cline{2-7} 
		&~\cite{7470935}                   &  ZigBee 802.145.4   &  Below 20m   &     -             &      Two Hexacopters   &   Measuring path loss exponent    \\ \hline
	\end{tabular}
\end{table*}

\begin{table*}[]
	\centering
	\caption{Simulation studies for channel modeling}
	\label{tbl:channel_simul}
	\begin{tabular}{|m{0.7cm}|m{0.7cm}|m{2.5cm}|m{1.5cm}|m{1.5cm}|m{3cm}|m{3.5cm}|}
		\hline
		\textbf{Type}                 & \textbf{Ref.} & \textbf{Frequency} & \textbf{Altitude} & \textbf{Environment} & \textbf{Simulation Details} & \textbf{Description} \\ \hline
		\multirow{3}{*}{\textbf{A2G}} & ~\cite{7037248}  &700MHz, 2000 MHz and 5800MHz      & 200m-3000m    &  Suburban, Urban, Dense and Highrise Urban   &  MATLAB, Wireless InSite, One quasi-stationary UAV                &    Finding a generic path loss model based on urban parameters           \\ \cline{2-7} 
		&     ~\cite{4151547}      &  5GHz    &    100m, 200m, 500m, 1000m and 2000m & Bristol area with irregular street    &      Ray tracing software            &    Modeling LOS probability  based on  building geometry                 \\ \hline
		\multirow{3}{*}{\textbf{A2A}} &   ~\cite{6167508}          &   --   & --    & --    &   Two UAVs (22m/s) in the simulated area       &       Modeling the packet dropout using Rician channel model        \\ \cline{2-7} 
		&    ~\cite{5700253}         & 2.4GHz (80MHz BW)      &  --   & --    &        Three UAVs, one as a relay          &   Modeling the bit rate over different channel models            \\  \hline
		\multirow{2}{*}{\textbf{G2A}} &~\cite{5267877}&  5GHz    &   --  &  --    &  MATLAB, one aircraft (300m/s)              &  Studying the characters of channel such as
		the Doppler, and the type of fading     \\ \hline
	\end{tabular}
\end{table*}

\vspace*{0.2cm}
{\revise{\emph{iii) 3GPP Study Item: UAV Performance Analysis}}}
\vspace*{0.2cm}
\label{sec:interferenceChallenges}

Based on the above traffic and channel characterizations, 
the companies involved in the SI evaluated the performance of cellular networks serving both airborne and ground users (GUEs). 
These studies demonstrate that UAVs are more likely to undergo downlink and uplink interference problems than GUEs~\cite{3GPP36777, GerGarUAVICC2018, GarGerGal_WCM_2018, AzaRosChi2017, AzaRosPol2017, 8647634, Yin2018UAV_GC, Ding2017varFactors, Ding2017UDNmag_IEEE}. 
This is mainly due to two factors, 
namely, that flying UAVs are likely to be in line-of-sight with a large number of base stations (BSs)~\cite{our_GC_paper_2015_HPPP}, 
and that the majority of these BSs are downtilted~\cite{Yang2018antDowntilt}, 
since their deployment has been optimized for providing coverage to GUEs. 
This impacts all phases of the cellular communication, 
i.e., 1) association and handover, 2) downlink transmissions, and 3) uplink transmissions:

\begin{figure}[!t]
\centering
\includegraphics[width=\columnwidth]{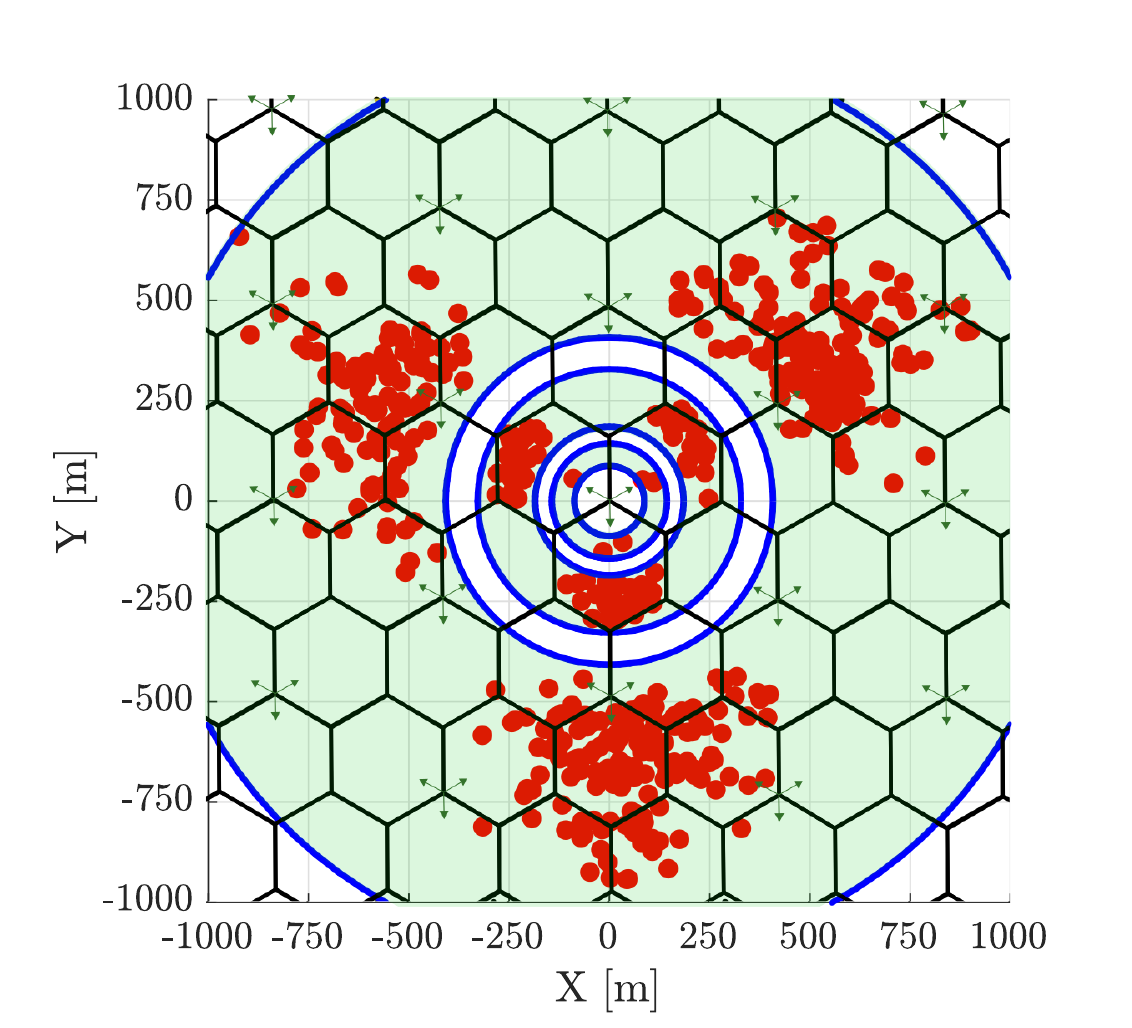}
\caption{2D location of 150\,meter-high UAVs (red dots) associated to a three-sector BS site located at the origin~\cite{GerGarUAVICC2018}. Hexagons illustrate the ground sectors covered by each BS.}
\label{fig:associationRegions}
\end{figure}

\begin{enumerate}
\item \emph{Association and handover:} In contrast to GUEs, 
flying UAVs do not generally associate to their physically closest BS. This is because cellular BSs generally focus their main antenna beam towards the center of their ground coverage area~\cite{Yang2018antDowntilt}. 
Instead, the association of airborne devices is dominated by the sidelobes of their directive BS antennas~\cite{GerGarUAVICC2018, 3GPP36777}. 
This can be observed in Fig.~\ref{fig:associationRegions}, 
which adopts the perspective of a tri-sector BS site located in the center of the scenario and illustrates the 2D location of its associated 150 meter-high UAVs (represented by red dots). 
These UAVs are clustered in three angular regions, 
which are consistent with the orientations of the central BSs ($120^{\circ}$, $240^{\circ}$, and $270^{\circ}$). 
Different association ranges highlighted in green can be observed, each corresponding to different antenna sidelobes of the $12^{\circ}$-downtilted BSs. 
Instead, the white regions delimit the areas where the antenna gain of the considered BSs is smaller than -30 dB, i.e., close to the radiation nulls of the downtilted antennas. 
In these areas, UAVs associate to BSs other than those located in the origin because they perceive an insufficient signal strength from the latter. 
As a result of the existence of these non-contiguous association regions and the reception of high-power interfering reference signals, 
UAVs experience increased outage and handover failure probabilities when compared to ground devices. 

\item \emph{Downlink transmissions:} As illustrated in Fig.~\ref{fig:interferenceProblem}, 
a UAV receives line-of-sight transmissions from a large number of BSs when increasing its altitude. 
Indeed, measurement campaigns have demonstrated that airborne devices flying at around 100 meters can receive signals from BSs located up to 10 kilometers away~\cite{AmorimWCOML2017}. 
This entails that a given UAV can undergo a substantial amount of interference from a multiplicity of ground BSs that transmit towards other GUEs or UAVs. 
Consequently, downlink transmissions towards aerial devices generally suffer from poor signal-to-interference-plus-noise ratios (SINRs) more often than their ground counterparts.

\begin{figure}[!t]
\centering
\includegraphics[width=0.98\columnwidth]{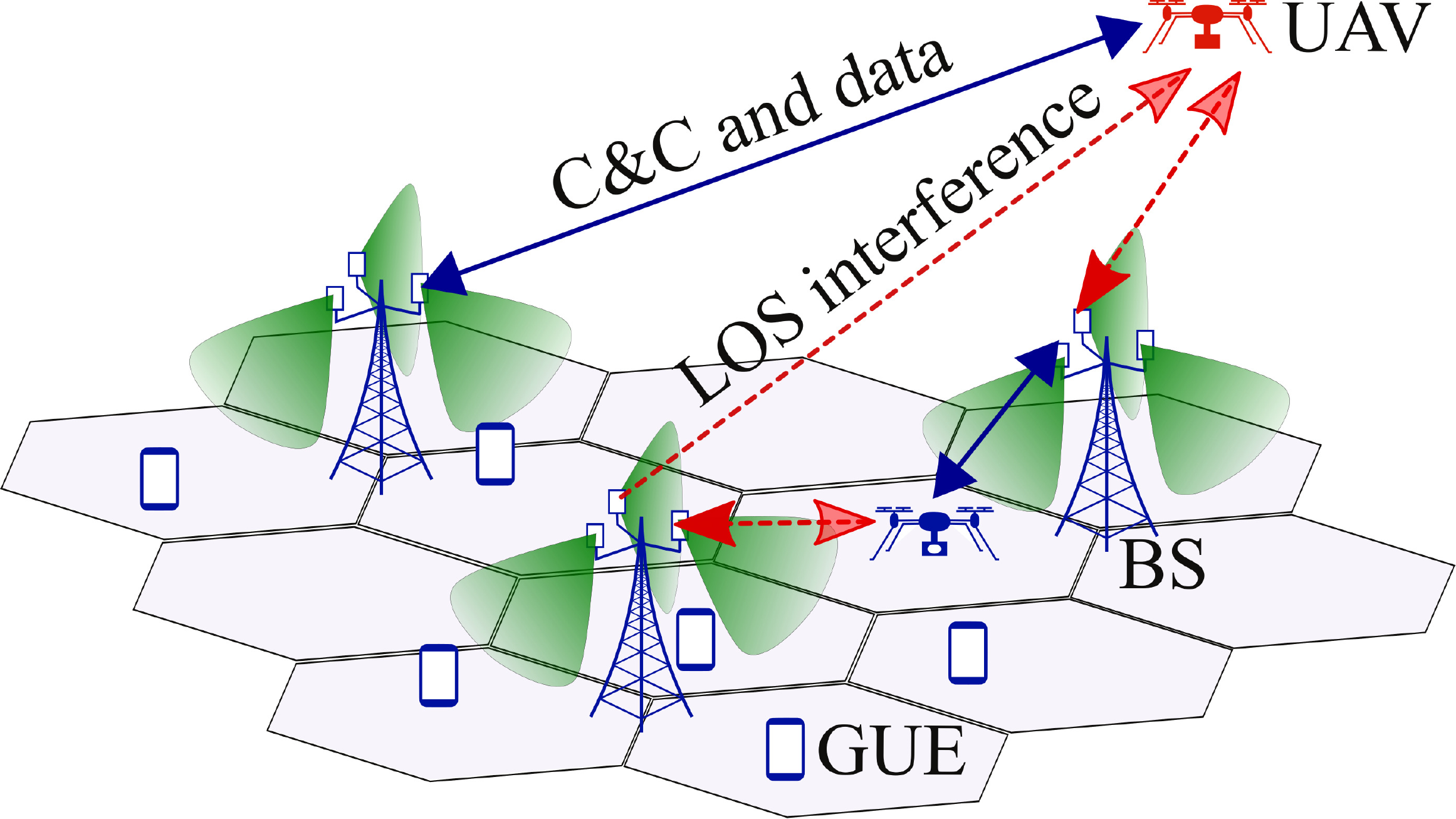}
\caption{Illustration of the UAV interference challenge in cellular networks: UAVs flying above the building clutter (e.g., the red-coloured UAV) generate/perceive interference towards/from a multiplicity of line-of-sight BSs~\cite{GerGarUAVICC2018, GarGerGal_WCM_2018}. Instead, UAVs flying at low altitudes (e.g., the blue-coloured UAV) only generate/perceive interference towards/from nearby BSs.}
\label{fig:interferenceProblem}
\end{figure}

\item \emph{Uplink transmissions:} UAVs' good propagation conditions with a multiplicity of BSs also impact the network's uplink performance. 
This is because airborne devices transmitting data towards their serving BS can generate strong interference to a variety of ground BSs, 
an issue that becomes critical in networks with a large number of UAVs~\cite{Li2018UAV_INISCOM}. 
As shown in Fig.~\ref{fig:interferenceProblem}, 
this UAV-generated interference has the potential of damaging the uplink communication of existing GUEs, 
who are more prone to have non-line-of-sight links with their serving BSs. 
Overall, both aerial and ground devices have been shown to experience diminished SINRs in networks with a substantial number of UAVs.

\end{enumerate}

\vspace*{0.2cm}
{\revise{\emph{iv) 3GPP Study Item: Enhancing UAV Communications}}}
\vspace*{0.2cm}
\label{sec:3GPPEnhancing}

To address the interference challenges described in Sec.~\ref{sec:interferenceChallenges}, the 3GPP examined a number of complementary interference mitigation techniques:

\begin{enumerate}
\item \emph{Association and handover:}
\begin{itemize}
\item \emph{UAV location and flight plan knowledge} can be leveraged to facilitate the handover procedure, e.g., by anticipating BS candidates for a potential handover. 
\item \emph{Enhancement of existing report mechanisms} through the definition of UAV-specific handover triggering conditions and an optimized control of the reporting load.
\end{itemize}
A detailed evaluation of these UAV mobility enhancement techniques is part of the work item (WI) study defined in~\cite{RP172826}, as detailed in Sec.~\ref{sec:WIPhase}. 
\item \emph{Downlink transmissions:}
\begin{itemize}
\item \emph{The full dimension MIMO (FD-MIMO)} multi-antenna BSs defined in LTE Release 13 enhance the performance of UAV communications thanks to a) their beamformed transmissions, which allow reducing the amount of interference generated towards the constrained spatial regions where UAVs lie, and b) their spatial multiplexing capabilities, which in turn enable a better utilization of the precious time/frequency resources.
\item \emph{UAVs with directional antennas and beamforming capabilities} contribute to reduce the number of downlink interferers perceived by aerial devices. 
These interference mitigation gains can be further complemented with a boost of the useful signal power in UAVs beamsteering towards their serving BS. 
Clearly, this solution entails a complexity increase in the design of hardware UAV transceivers.
\item \emph{Cooperative multipoint (CoMP)} can convert the harmful line-of-sight BS interferers into useful signal contributors. Indeed, both UAV control and data transmissions benefit from intra- and inter-site cooperation~\cite{Book_CoMP}. 
However, the gains attainable with this approach are limited in practical deployments due to the large number of BSs interfering towards a given UAV and the increased inter-BS signalling load. 
\end{itemize}
\item \emph{Uplink transmissions:}
\begin{itemize}
\item \emph{Uplink power control} is essential to harmonize the coexistence among GUEs and aerial devices. 
In this line, approaches that define different fractional path loss compensation factors and offsets $P_{0}$ for UAVs and GUEs proved effective in mitigating the interference generated by UAVs~\cite{Ding2017ULlos}.
\item \emph{Full dimension MIMO (FD-MIMO)} can also benefit uplink transmissions by enabling spatial separation of ground and aerial users, which have clearly distinguishable propagation characteristics.
\item \emph{UAVs with directional antennas and beamforming capabilities} generate a diminished amount of interference to the GUE-generated uplink transmissions owing to a potential decrease in both the number of interfered BSs and the UAV transmission power.
\end{itemize}
\end{enumerate}

As the reader might have noticed, 
updated information about the flying status of a mobile device is required to effectively implement some of the above-mentioned solutions. 
The 3GPP considered a variety of device- and network-based solutions to acquire this information. 
Among others, 
these include the use of explicit UAV identification signalling, 
the exploitation of mobility history information, 
or the employment of measurement reports from mobile devices.

\subsection{\revise{3GPP Work Item: Main Outcomes and Way Forward}}
\label{sec:WIPhase}

The above-mentioned observations led to the definition of a WI in December 2017~\cite{RP172826}. {\revise{The core part of this WI was concluded in September 2018 and it provides further enhancements to LTE in the following areas~\cite{3gpp_RP_181644}:}}
\begin{itemize}
	\item \emph{Uplink power control}, {\revise{by augmenting the existing fractional power control mechanisms through a) the assignment of a UAV-specific path loss compensation factor, and b) the range extension of the $P_{0}$ parameter~\cite{3gpp_TS36331}.}}
	\item \emph{Signaling}, to identify the status of airborne devices. {\revise{For instance, the technical specification (TS) 36.331 now defines new reporting events that are triggered when the UAV height is above or below a BS-configurable threshold~\cite{3gpp_TS36331}.}}
	\item \emph{Interference detection}, {\revise{by allowing UEs to trigger a measurement report when a configurable number of reference signals received from neighboring cells satisfy specific power-related conditions~\cite{3gpp_TS36331}.}}
	\item \emph{Subscription-based access}, {\revise{to prevent the non-authorized connection of cellular-connected UAVs. This is enabled through new network signaling from the core network to cellular BSs, allowing the latter to determine whether the UE subscription includes  aerial functionalities.}}
	\item \emph{Mobility}, {\revise{where new radio resource control (RRC) signalling was included to facilitate the flight plan communication from UAVs to their serving BS. This information can be subsequently exploited to facilitate handovers.}}
\end{itemize}

Ultimately, it is expected that the 3GPP continues its work on the UAV space in the future, possibly through the definition of a new WI focused on 5G-based solutions like massive MIMO~\cite{GerGarUAVICC2018, GerGarGal2018, GarGerGal_WCM_2018, ChaLar2017, ChaDanLar2017}.

\subsection{\revise{UAV Standardization Outside the 3GPP}}

{\revise{Other standardization bodies have also considered the particular characteristics of UAVs in the definition of new specifications:}}
\begin{itemize}
	\item {\revise{\emph{The International Telecommunication Union Telecommunication Standardization Sector (ITU-T)} defined the work item Y.UAV.arch for providing a functional architecture for UAVs and UAV controllers using IMT-2020 networks~\cite{ITU_WI}. 
Additionally, this work item also aims at defining capabilities in the application layer, 
the service and application support layer, 
as well as implementing security measures for facilitating the integration of UAVs in IMT-2020 networks.}}
	\item {\revise{\emph{The European Telecommunications Standards Institute (ETSI)} technical report (TR) 103 373 aims at identifying UAV-specific use cases and understanding whether new spectrum rules are required for enabling them~\cite{ETSI_TR36331}. 
Additionally, 
the specifications of UAV wireless communications have been pondered for determining how the future Internet Protocol (IP) suite architecture should be shaped~\cite{ETSI_GSNGP001}.}}
	\item {\revise{\emph{The Institute of Electrical and Electronics Engineers (IEEE)} defined the Drones Working Group in 2015. 
This group aims to develop a standard for consumer drones, primarily with the intention of addressing privacy and security concerns. 
With this purpose, 
the standard is currently focused on specifying 1) the taxonomy and definitions related to UAVs~\cite{IEEE_P20251}, 
and 2) the requirements, systems, methods, testing and verification required to preserve the privacy and security of people and properties within range of the UAVs~\cite{IEEE_P20252}.}}
\end{itemize}

%

%% file: AerialBaseStations.tex
\graphicspath{{Figures/}}

\section{Aerial Base Stations: Challenges and Opportunities}
\label{s:flying}

The falling cost and increasing sophistication of consumer UAVs combined with miniaturization of BS electronics have made it technically feasible to deploy BSs on flying UAVs. 
Because UAV BSs can be quickly deployed at optimum locations in 3D space, 
they can potentially provide much better performance in terms of coverage, load balancing, spectral efficiency, and user experience compared to existing ground-based solutions. 

The deployment of aerial BSs, however, faces several practical issues. 
In particular, placement and mobility optimization are challenging problems for aerial BSs, 
which have attracted significant attention from the research community. 
The optimization of UAV power consumption and the development of practical recharging solutions for UAVs are also important challenges to overcome for sustaining the operation of aerial BSs. 
{\revise{Finally, the optimization of the end to end link when a moving UAV connects a ground user to the wired backbone of the network is a non-trivial problem.}} 
In this section, we survey recent research addressing these issues.


\subsection{Placement Optimization for Aerial BSs} 
{\revise The problem of optimum placement is more challenging for aerial BSs compared to the conventional terrestrial BSs because an aerial BS can be placed at many different heights in the sky~\cite{8647634, Liu2018UAV_ICCWS}. 
However, the coverage as well as the UL and DL channels change with the altitude of the BS. 
Different researchers employed different algorithms to solve the placement optimization problem for aerial BSs. 
Some researchers considered the height of the aerial BS as a variable in their optimization formulation thus treating it as a 3D placement problem, 
while others essentially solved 2D placement problems for constant heights. 
Optimizations also differed in whether the backhaul, 
interference from other BSs, 
and existence of terrestrial BSs in the same coverage area were considered in problem formulation. 
Finally, 
the objectives in of these optimizations varied from maximizing the system capacity to minimizing the required number of aerial BSs to minimizing the total transmit power of the entire aerial system. 
In this section, 
we survey these optimization studies  and compare their main features in Table~\ref{t:place}.
	
Using brute force search to maximize the 5th percentile spectral efficiency of the system, 
Merwaday et al.,~\cite{7122576} optimized 2D placements of a small number of `helper' aerial BSs to improve coverage of a large service area where some of the terrestrial BSs are damaged by natural disasters or malicious attacks. 
Not surprisingly, 
the optimal locations for the aerial BSs were proposed to be along the cell edges, 
which has the most effect on the 5th percentile throughput. 
A less intuitive observation was that the height had little effect on the overall throughput, 
but lowering the height of the aerial BSs markedly improved 5th percentile throughput. 
This was attributed to rather line-of-sight environments in the simulations, 
which caused better SNR for cell-edge users due to smaller distance to the aerial BSs. 
Similar 2D placement optimizations were also studied by Rohde et. al.,~\cite{Rohde20131893} and Galkin et. al.,~\cite{7461487}, 
but instead of brute force search, they employed genetic algorithm and K-means clustering, respectively, to solve the optimization problem.  
	
Yaliniz et. al.,~\cite{bor2016efficient} attempted 3D placement optimization for aerial BSs in the context of heterogeneous networks (aerial BSs are used to augment terrestrial BSs), 
but observed that there are quadratic, exponential, and binary terms in the problem formulation, 
which makes it a mixed integer non-linear problem. 
The authors show that the problem can be efficiently solved by using a combination of the interior point optimizer and bisection search. 
Mozaffari et. al.,~\cite{7510870} also studied 3D placement optimizations, 
but sought to minimize the total transmit power of a homogeneous network of aerial BSs. 
For such scenarios, 
the authors solved the 3D placement optimization problem by dividing the problem into two sub-problems that are solved iteratively. 
Given the height of the aerial BSs, 
the first sub-problem obtains the optimal locations of the UAVs using the \textit{facility location framework}. 
In the second sub-problem, 
the locations of aerial base stations are assumed to be fixed, 
and the optimal heights are obtained using tools from \textit{optimal transport theory}.
	
The authors of~\cite{7122576,Rohde20131893,7461487,bor2016efficient,7510870} sought to maximize the system performance or minimize the total transmit power of the network.
In contrast, 
Ko\v{s}merl et. al.,~\cite{6881196} and Lyu et. al., ~\cite{7762053} analyzed the placement optimization problem with the objective of obtaining the minimum number of aerial BSs that can meet target user requirements in a given service area.
While Ko\v{s}merl et. al.,~\cite{6881196} employed evolutionary computing to solve the problem, 
the authors of ~\cite{7762053} observed that it can be formulated as a Geometric Disk Cover (GDC) problem, 
whose objective is to cover a set of ground users in a region with the minimum number of disks of given radii where the radii is influenced by the height of the UAVs. 
Since the GDC problem is NP-hard, 
a polynomial-time heuristic, 
called spiral placement algorithm, 
was proposed to solve the problem approximately. 
It was shown that the proposed spiral placement algorithm achieved near optimal solutions with significantly lower complexity compared to the optimal solution.  
	
When a ground user is associated with an aerial BS, 
data is downloaded or uploaded using two distinct wireless links, 
the access link that connects the user to the UAV and the backhaul link which connects the UAV to the core network. 
The placement of the UAV will affect both of these links. 
Works in~\cite{7122576,Rohde20131893,7461487,bor2016efficient,7510870,6881196,7762053} considers only the access link in their optimization formulation and assumes that the backhaul link has infinite bandwidth. 
In a recent work, Sun and Ansari~\cite{ansari_ICC2018} optimized the vertical placement of the aerial base station by jointly considering the spectral efficiency of the backhaul link as well as the access link.}

%
%

\begin{table*}[!t]
	\caption{\revise{Comparison of aerial BS placement algorithms}}
	\label{t:place}  
	\begin{center}
		\renewcommand{\arraystretch}{1.5}
		\begin{tabu} to \hsize {|l|X|c|c|X|}
			\hline
			Reference & Objective & Dimension & Backhaul & Algorithm \\ 
			\hline
			Merwaday et. al.,~\cite{7122576} &  Maximize 5th percentile throughput  &  2D & Ignored & Brute force search \\
			\hline
			Rohde et. al.,~\cite{Rohde20131893} & Maximize downlink throughput & 2D & Ignored & Genetic algorithm \\
			\hline
			Galkin et. al.,~\cite{7461487} & Maximize downlink RSS & 2D & Ignored & K-means clustering \\
			\hline
			Yaliniz et. al.,~\cite{bor2016efficient} &  Maximize network revenue & 3D & Ignored & Combination of \textit{interior point optimizer} and \textit{bisection search} \\
			\hline
			Mozaffari et. al.,~\cite{7510870} &  Minimize total transmit power & 3D & Ignored & facility location and optimal transport theory  \\
			\hline
			Ko\v{s}merl et. al.,~\cite{6881196}    &   Minimize required number of aerial BSs   & 2D & Ignored & Evolutionary Computing \\
			\hline
			Lyu et. al., ~\cite{7762053}     &   Minimize required number of aerial BSs  & 2D & Ignored & Spiral placement algorithm \\
			\hline	
			Sun and Ansari~\cite{ansari_ICC2018}     &   Maximize spectral efficiency  &  1D & Considered & STABLE (self-derived algorithm) \\
			\hline		
		\end{tabu}
		1D: optimize vertical placement, i.e., the height for a specific 2D ground location \\
		2D: optimize $x$ and $y$ coordinates on a horizontal plane at a specific height \\
		3D: optimize horizontal coordinates as well as the height 
	\end{center}
\end{table*}


\subsection{Mobility Optimization for Aerial BSs} 

Placement optimization studies surveyed in Table~\ref{t:place} do not include mobility of aerial BSs in problem formulation. 
Mobility, however, is an intrinsic feature and capability of aerial BSs, 
which provide additional opportunities to dynamically improve their placements in response to user movements on the ground. 
To exploit the mobility features of aerial BSs, 
the practical hardware limitations of the UAVs must be considered. 
The limitations on speed and accelerations are studied through filed experiments in~\cite{7974336} using a consumer UAV.
Based on the transportation methods of aerial BSs, 
there are two types of mobilities considered in the literature:
\begin{enumerate}
	\item UAVs are used only to transport a BS to a particular ground location where the BS autostarts to serve the users.
If the BS needs to be relocated, it must shut down first before being transported to the new location. 
This type of aerial BSs  therefore cannot serve while it is in motion, 
but it can resume its service as soon as it reaches a target location. 
As will be explained later in Section~\ref{subs:nokiaprototype}, 
prototypes from Nokia Bell Labs~\cite{nokiafcell} fall into this category. 
	\item UAVs continue to carry the BSs and the BSs can continuously serve the ground users while they are flying. 
For example, 
prototypes from Eurecom~\cite{PerfumeProject}, 
later described in Section~\ref{subs:eurecomprototype}, 
belong to this category.  
\end{enumerate}

Considering the first type of aerial BSs, 
Chou et. al.,~\cite{7037574} studied a BS placement mechanism where the ground users are not served by the BS when it is moving to a new location. 
The loss of service time due to BS mobility therefore becomes a critical parameter for the optimization. 
An aerial BS in this case should consider both the user density of the target location as well as the  moving time to the new location when deciding its target location. 
The authors of~\cite{7037574} have shown that this problem can be modelled as \textit{facility location problem}, 
where the transport cost represents the loss of service time due to the movement of the BS from previous location to the new location.

The second type of aerial BS mobility opens up new opportunities to employ aerial BSs due to their ability to serve ground users while in motion.
In particular, 
under this scenario,
the cost of BS mobility becomes negligible. 
It is then possible to design more advanced solutions where aerial BSs can continuously cruise the service area to maximise network performance under geospatial variance of demands.

{\revise{In both types, although a single UAV can perform plenty of tasks, multiple UAVs can form a cooperative group to achieve an objective more efficiently, and to increase the chance of successful task operation. Additionally, the robustness of the communications will increase by employing cooperative UAVs~\cite{8528342}. Maintaining the connectivity and controlling the distance between multiple UAVs is one of the main challenges in employing cooperative UAVs~\cite{7990240}. Maximizing the coverage area~\cite{8485481}, cooperative carrying task~\cite{8407124, 8444215}, searching and localizing a target~\cite{7520118} are among the tasks that can be done by multiple UAVs.}}

Designing cruising aerial BSs requires autonomous mobility control algorithms that can continuously adjust the movement direction or heading of the BS in a way that maximizes system performance. 
These algorithms must also insure that multiple aerial BSs cruising in an area can maintain a safe distance from each other to avoid collisions.  
Fotouhi et. al.,~\cite{7974285} proposed distributed algorithms that take the interference signals, 
mobile users' locations and the received signal strengths at UEs into consideration to find the best direction for BS movements at any time. 
Controlling the mobility of a single serving UAV is also discussed in~\cite{ourwork_gc16}. 
Game theoretic mobility control algorithms are proposed in~\cite{our_ieeeaccess, ourwork_gc17} for multiple aerial base stations cruising freely over a large service area without being subject to individual geofencing.
The game theoretic mobility control not only increased 5th-percentile packet throughput by 4x compared to hovering BSs, 
it also helped avoid collisions as the BSs were implicitly motivated to move towards different directions to maximize coverage and throughput.  
{\revise{The trajectory of a single UAV also can be optimized to improve the system performance. 
A UAV with a mission to fly between a source and destination point is studied in~\cite{p18Zhang}. 
During this mission it has to maintain a reliable connection by associating with a  ground base stations at each time. 
In~\cite{8329013} a UAV is used to offload data traffic from cell edge users and improve their performance. 
It is shown that by using one single UAV and optimizing its trajectory, 
the throughput improves significantly compared with the conventional cell-edge throughput enhancement scheme with multiple micro/small cells.
		
}}

\subsection{Power-Efficient Aerial BSs}  

A critical problem of Aerial BSs is their short lifetime due to the battery depletion problem. 
Power-efficient operation therefore is must to extend the battery lifetime. Power is consumed by both communications (electronics) and mobility (mechanical). 
Researchers therefore worked on both types of energy saving.



\subsection*{Reducing Communication Energy} \label{subsub:reduce_comm_eng}

One alternative to reduce the communication energy of UAV base stations is to minimize the \textit{transmission power}.
For example, 
minimizing the transmission power of one UAV~\cite{7417609} or multiple UAVs~\cite{7486987} when they are deployed in the optimal location to cover the target area is studied in the literature.
The deployment of UAVs was optimized to minimize the total transmit power for UAVs while satisfying the users' data-rate requirements is discussed in~\cite{6781225,7510870}.
Reducing the number of transmissions to decrease the energy consumption of a UAV is addressed in~\cite{7412759}. 
In this work, the minimum number of stopping and transmission points for a UAV to cover all downlink users are derived.

Another solution to improve communication energy efficiency is to develop optimal \textit{transmission schedule} of UAVs, 
especially when UAVs are flying in a predetermined trajectory.
{\revise{A scenario where UAVs are employed to collect sensor data and forward them to a remote base station is studied in~\cite{7192644}. 
The frequent need of UAVs to recharge interrupts the data collection, as a result, prolonging the UAVs lifetime is critical. 
It is assumed that both UAVs and sensors exploit TDMA for data transmission. 
In each TDMA frame, sensors broadcast packets to all UAVs, 
and UAVs report the reception qualities to the remote BS. In return, the BS proposes a scheduling model to minimize the energy consumption of UAVs while guaranteeing the required quality. 
A sub-optimal algorithm is developed for indicating the allocation of packets, 
time slot and power for each UAV. 
The simulation results show that the proposed algorithm can extend UAVs lifetime around 60\% compared with some existing packet allocation algorithms.}}
{\revise{An energy efficient UAV system is addressed in~\cite{6848000} where a UAV is flying in a circular trajectory over a sensor field to collect data from sensor clusters. 
Apparently, the movement of UAV results in varying distances between cluster heads and the UAV, which consequently influences on the quality of data transmission and energy efficiency. 
To improve the energy efficiency, 
a game theoretic data collection method is proposed to optimize the allocation of time slots to cluster heads to send data to the UAV. 
The cluster heads are the intelligent players, and the utility function reflects the number of transmitted bits over the amount of consumed energy. 
This work is extended for multiple UAVs in~\cite{6867394}.
}}


Optimal scheduling for beaconing messages in order to maximize the energy efficiency of two competing UAVs is discussed in~\cite{7470936}. 
In this work, two UAVs are moving randomly over areas including mobile users, and send periodically beaconing messages to users to announce their presence. 
A non-cooperative game theory is proposed for finding the best beaconing period for UAVs.
Energy efficient uplink transmission between a terrestrial link and single LAP aerial destination is addressed in~\cite{6978873}. 
Terrestrial nodes can select either to communicate directly with LAP or to use other terrestrial nodes as relay to reduce energy cost. 
Context aware network is assumed in this work, where nodes are aware of the necessary transmission and channel parameters (through gossip control).

Although the proposed methods successfully reduce the communication energy consumption for UAVs, 
one disadvantage of these alternatives is that the ratio of communication energy consumption to the total energy consumption of UAVs is generally negligible~\cite{ZORBAS201380}. 
This observation motivates the efforts in reducing the mechanical energy consumed by UAVs, as detailed in the following.

\subsection*{Reducing Mechanical Energy}

To reduce mechanical energy of UAVs, first, an energy consumption model is needed.
According to~\cite{DiPugliaPugliese2016,Zorbas201616}, the energy consumption of UAV can be modeled by 
\begin{equation}
E = (\beta+\alpha.h)t + P_{max}(h/v),
\end{equation}
where
$\beta$ is the minimum power needed to hover just over the ground (when altitude is almost zero),
and $\alpha$ is a motor speed multiplier. 
Both $\beta$ and $\alpha$ depend on the weight and motor/propeller characteristics. 
$P_{max}$ is the maximum power of the motor, $v$ is the speed, 
and $t$ is the operating time. 
The term $P_{max}(v/h)$ refers to the power consumption needed to lift to height $h$ with speed $v$.

Another model for energy consumption is defined in~\cite{ZORBAS201380} where the energy consumption of drone is related to its altitude. 
Following this work, 
energy consumption of drone can be calculated by $m \cdot g \cdot h$, 
where $m$ is the mass of the drone,
$g$ is the gravitational acceleration, 
and $h$ is the altitude of the drone.

According to these models, 
one solution to control the energy consumption of UAVs is to regulate their height. 
However, changing the height might reduce the performance of UAVs. 
For example, in target coverage, there is a tradeoff between energy consumption and coverage radius.
Higher altitude means higher observation radius but higher energy consumption~\cite{DiPugliaPugliese2016,Zorbas201616,ZORBAS201380}. 
Optimizing the flight radius and speed to improve energy efficiency is also addressed in~\cite{7888557}. 
One of the major advantages of these methods is that they have targeted the mechanical energy consumption of UAVs, 
which is considered as the main source of energy consumption for UAVs.

Given that mechanical activities consume much more power compared to electrical activities,
manoeuvring of UAV BSs must be controlled in a power-efficient manner. 
Algorithms that consider the battery and energy consumption of UAVs as a constraint have also been studied in the literature. 
For instance, some works consider a limited availability of energy~\cite{Yoo2016140,7101619,Torres2016441}, 
and a limited flight time~\cite{5928811} in developing path planning algorithms.

\subsection{Recharging of Aerial BSs}
Separate from reducing the energy consumption of UAVs, 
another attractive solution to combat the short lifetime of UAVs is to consider charging locations for them and replace exhausted UAVs with the fully charged ones. 
Comparing to the methods that focus on reducing the energy consumption, 
this solution is more costly and complex, as replacing/charging points must be designed in urban areas~\cite{Sharma2016,DiPugliaPugliese2016,dronechargemelbourne}. 
Moreover, the battery consumption of UAVs needs to be monitored regularly.

Sharma et al.~\cite{Sharma2016} propose monitoring the battery level of UAVs by Macro base stations (MBSs). 
When the battery reaches a critical value, 
they are returned to MBS, and already charged ones replace them.
Similarly, replacing UAVs by new ones is addressed in~\cite{DiPugliaPugliese2016}. 
In this work, several UAVs exist to monitor mobile targets; however, an optimization problem is formulated to minimize the number of required UAVs. 
An algorithm for automating the replacement of UAVs is presented in~\cite{ERDELJ2017}, 
which can be used in a multi-UAVs environment. 
Employing this algorithm provides continuous uninterrupted service for users.
Small recharging garages in BS towers, 
and in power-lines of urban areas  are proposed by~\cite{7994915} and~\cite{dronechargemelbourne}, respectively.

\subsection{\revise{Fronthauling and access communication links}}
{\revise{
		Assuming that it is feasible to efficiently fly a BS attached to a UAV in terms of load, 
dynamic positioning and power consumption, 
a crucial remaining aspect is the establishment of a communication link between the UAV and the terrestrial network. 
This link, conventionally referred to as fronthaul,
is crucial to guarantee enough bandwidth and reliability in the access link between the UAV and the user terminals.

Due to the need of keeping the computational complexity low, 
thus the associated energy consumption, 
it is foreseen aerial BS will generally work as relay nodes that requires the implementation of a reduced number of protocol stack layers, 
with the simplest configuration that involves only layer-1 and substantially works as an amplify and forward node.
Moreover, another crucial aspect to minimize the use of the spectrum will be the possibility for aerial BS to allow fronthaul and access communications on the same frequencies. 
Recent focus in the 3GPP 5G New Radio (NR) Release 15 was precisely on Integrated Access and Backhaul (IAB) network architectures~\cite{TR_38874}. 
		
Taking into considerations the above mentioned aspects,
the validation of the end-to-end performance of such a system constitutes a fundamental pillar for justifying the adoption of aerial BSs in cellular communication systems. 
		
Initial attempts towards this goal can be found in~\cite{2018BonfanteSBH_journal, 8424236_RelayUAV_Alouini, 2018arXiv180707230F}. 
In~\cite{2018BonfanteSBH_journal},
the authors analyse the achievable end-to-end system performance with different deployment of small cell relays served by the spatial multiplexing capabilities of a massive MIMO wireless fronthaul link. 
Although not explicitly addressing the UAV scenario, 
aerial BSs are indicated in the paper as an important upgrade to increase the required deployment flexibility to jointly maximize fronthaul and access propagation conditions. 
In line with this conclusion, 
and focusing more on the aerial BS case, in~\cite{8424236_RelayUAV_Alouini}, 
the authors study the use of multiple UAVs for multi-hop relaying communications. 
The placement of the UAVs is optimized to maximize the end-to-end signal-to-noise ratio when applying different relay schemes, 
namely decode-and-forward and amplify-and-forward. 
Additionally, in~\cite{2018arXiv180707230F} a multi-tier 5G scenario with IAB architecture, 
adopting massive MIMO terrestrial macro BSs and full-duplex (FD) drone small cells, is investigated. 
The achievable performance are presented for scenarios with only one terrestrial macro BS, 
and one or more hovering aerial BS. 
The obtained conclusions do not consider the potential interference generated from and towards neighbouring terrestrial macro BS which, 
due to high probability of LOS conditions of the aerial BS, 
might result in harmful impact on a large geographical area.    
}}

\graphicspath{{Figures/}}
\subsection{Paradigm-Shifting Cost Model of Aerial BSs}
\label{s:CostModel}

\begin{figure}
	\centering
	\subfigure[CAPEX]{
		\begin{minipage}[b]{0.5\textwidth}\includegraphics[width=\linewidth]{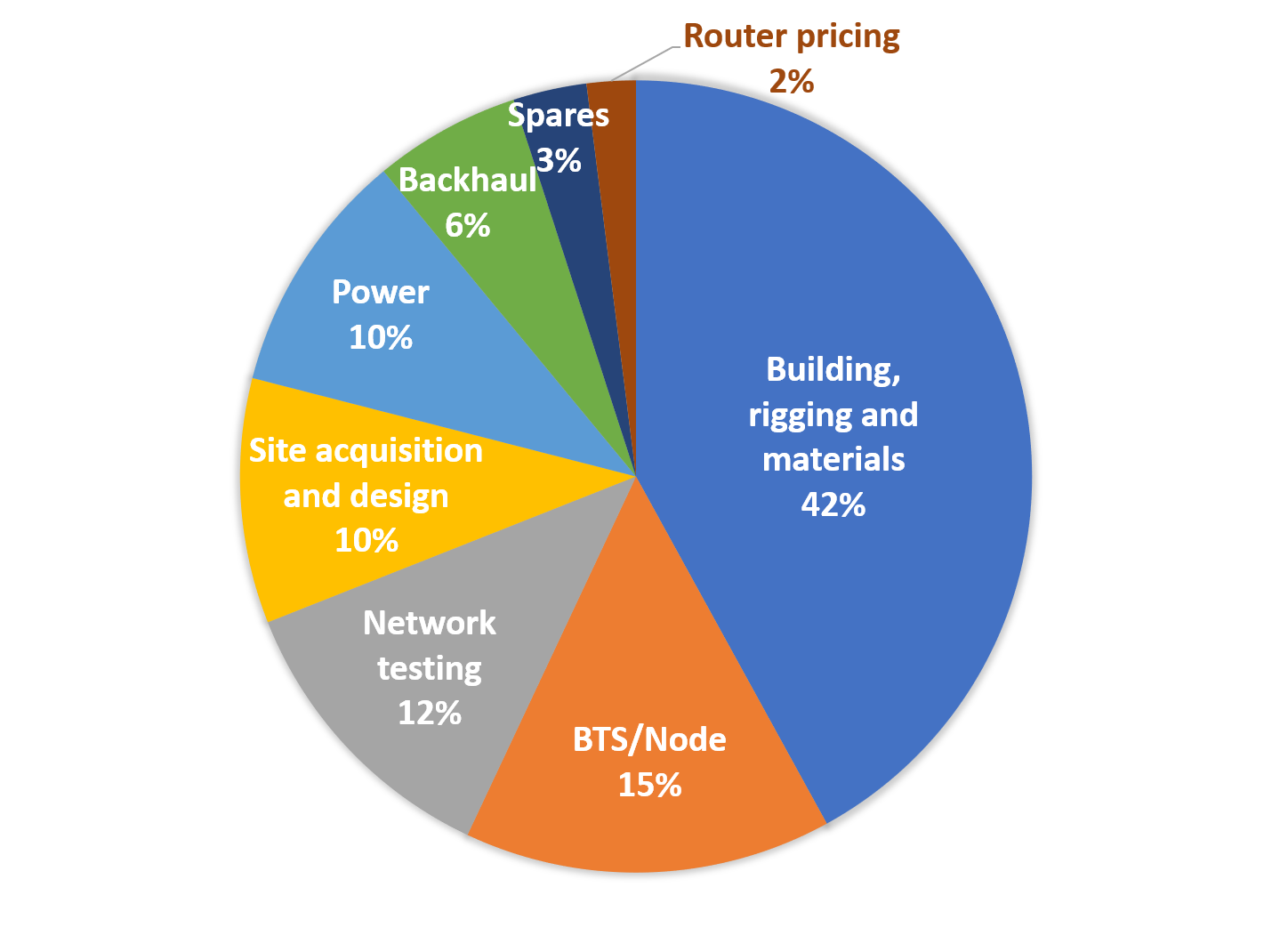}
		\end{minipage}
	}
	\subfigure[OPEX]{
		\begin{minipage}[b]{0.5\textwidth}\includegraphics[width=\linewidth]{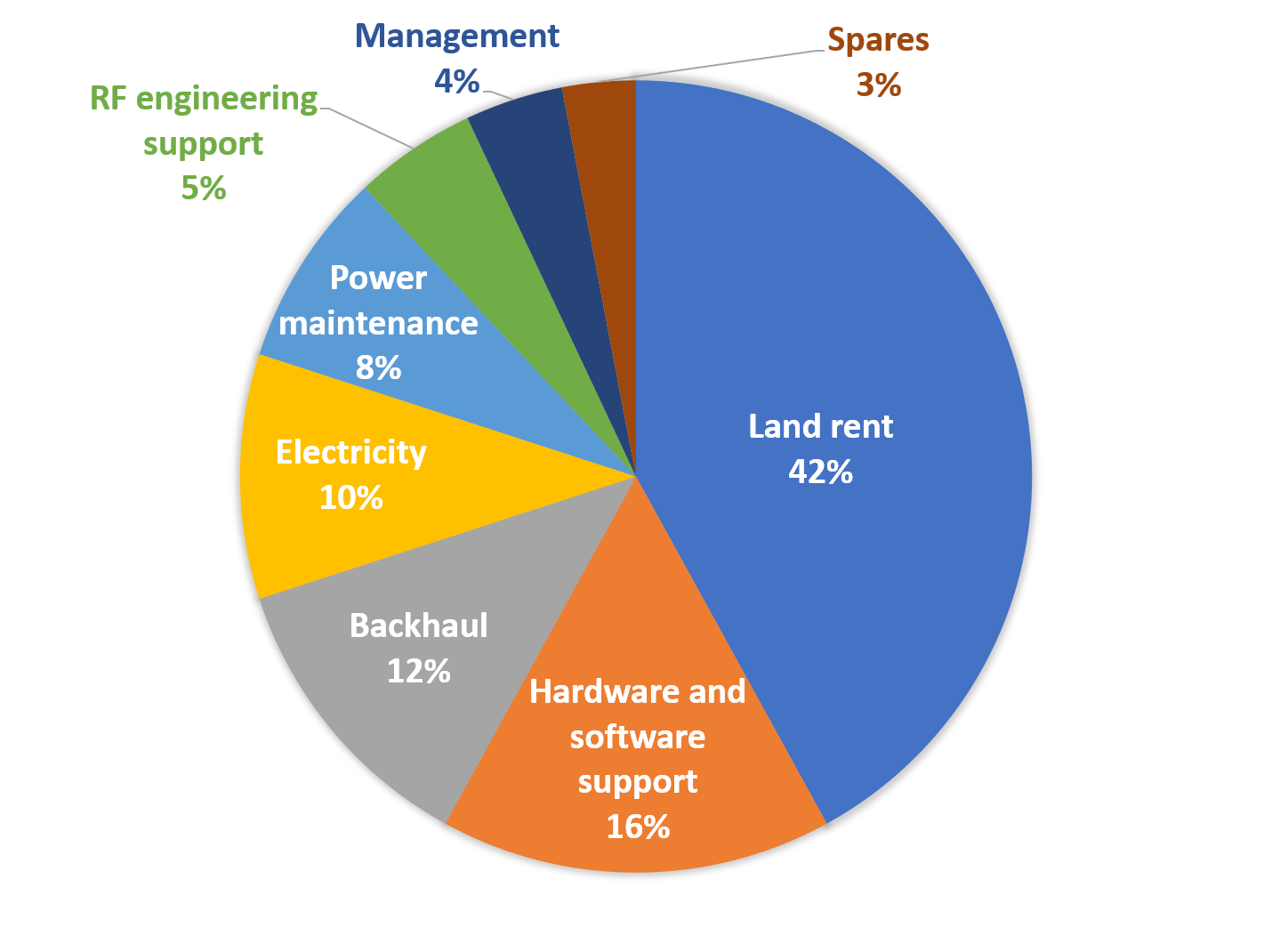}
		\end{minipage}
	}
	\caption{CAPEX/OPEX breakdown in developed countries~\cite{instance1290}.} 
	\label{fig: CAPEX/OPEX}
\end{figure}

Next generation cellular networks are expected to be 50 times more cost effective than 4G~\cite{roh20145g, 6867394}. 
Therefore, cost saving has become a major challenge for conventional cellular operators. 
Costs of conventional cellular operators usually include capital expenditure (CAPEX) and operational expenditure (OPEX). CAPEX comprises acquisition, 
design and construction of site, 
purchase and implementation of equipment, etc. 
OPEX is made up of recurring costs such as site maintenance and rental, 
personnel expenses, electricity, etc. 
Fig.~\ref{fig: CAPEX/OPEX} depicts the CAPEX/OPEX breakdown in developed countries.


It can be observed from Fig.~\ref{fig: CAPEX/OPEX} that cellular operators spend more than half (52\,\%) of their CAPEX on site acquisition and construction, 
which is followed by the costs of BSs, network testing, power and backhaul. 
However, CAPEX breakdown for drone-cells is expected to be quite different. 
With flying drones carrying cellular BSs, site-related costs could be significantly reduced or even completely removed. 
While the costs of BSs and network testing for drone-cells should increase compared to employment of terrestrial BSs due to the purchase and operation of drones. 
Overall, drone-mounted BSs have the potential to slash down the CAPEX of cellular operators by significantly reducing the site-related costs, 
which currently account for the lion's share of the CAPEX.

In terms of OPEX, site-related costs also occupy a major proportion. 
Site rental makes this percentage even larger, 
if cellular operators do not own their sites~\cite{meddour2011role}. 
As it can be seen from Fig.~\ref{fig: CAPEX/OPEX}, 
land rent constitutes up to 42\% of cellular operators' OPEX in developed countries. 
The authors in~\cite{lang2009business} also reveal that site rental dominates the OPEX of cellular operators. 
Moreover, due to site shortage and stricter environmental regulations, site rental is becoming increasingly expensive~\cite{RNSLTE}.
Furthermore, promising 5G techniques such as ultra-dense cell deployment make the costs of site rental even higher since a large number of BSs will be required. 
When using flying BSs, the most expensive part of OPEX could be greatly reduced, at the cost of potentially higher expenses on electricity and backhaul. 
There will also be some additional costs such as annual registration fee for drones, flying insurance and their battery replacement.

Compared to traditional cells equipped with terrestrial BSs, 
cost model of drone-cells is paradigm-shifting. With BSs mounted on flying drones, 
both CAPEX and OPEX of conventional cellular operators could be significantly reduced by saving site-related costs, 
which contributes to the improvement of cost effectiveness for conventional cellular operators.
In addition, when employing terrestrial BSs or relays to extend the cellular coverage, 
cell reorganization is required and could be expensive. 
This cost could be saved by deploying flying BSs/relays instead. 
Moreover, new business models may also emerge. 
For example, the deployment of drone BSs could follow the time-variant movement of people/cellular UEs, 
which could be achieved by easy and cheap shipping means such as public transportation systems (e.g. trains and buses). 
For another example, such paradigm-shifting cost model also enables some new cellular operators to provide opportunistic communication services. 
For drone fleet owners such as Google and Amazon, goods and data could be delivered simultaneously by flying drones following designated routes.

%% file: Prototyping.tex
\section{Prototyping and Field Tests} \label{s:DronePrototypingAndFieldTests}

Several drone communication prototypes have been already presented in literature or shown to a wider audience during exhibitions and commercial events. 
Their main scope is to extend coverage where ground wireless infrastructures are not feasible or to improve end user performance using flexible and dynamic deployment of serving base stations where required. 
In this section, we provide a description of the most interesting ones targeting both high altitude (Facebook Aquila and Google Loon) and low altitude (Nokia F-Cell and Eurecom Perfume) applications. 
We also discuss an example of `digital sky ecosystem' (Huawei) designed to promote and study use cases and applications. 
Finally, additional examples of testbeds involving drones are presented. 

\subsection{Facebook Aquila}
\begin{figure}[t]
	\includegraphics[scale=0.35]{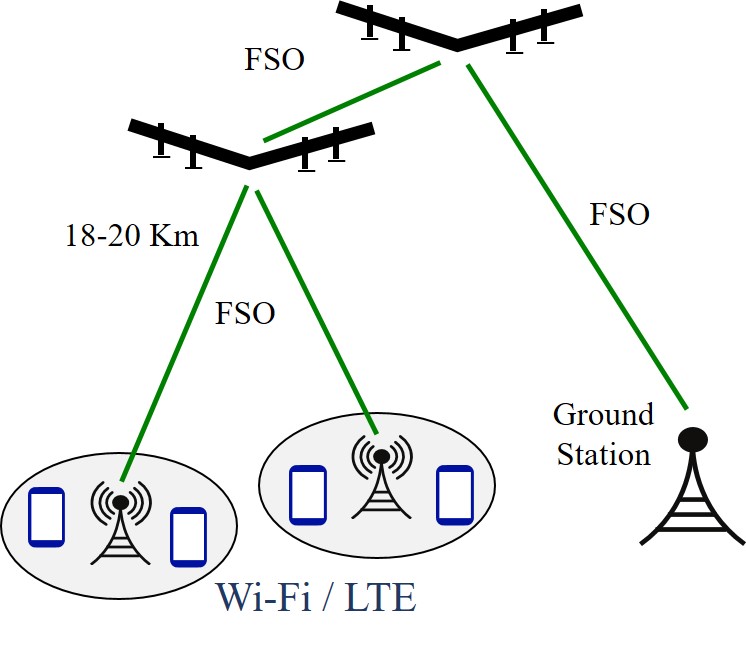}
	\centering
	\caption{Illustration of Facebook Aquila system architecture.}
	\label{fig:AquilaSketches}
\end{figure}

One interesting example of high altitude drone BS is the Facebook Aquila project~\cite{AquilaProject}~\cite{AquilaProjectSpectrumMag}, 
which aims at providing Internet coverage in remote areas directly from the sky. 
The main component of the system architecture is an unmanned autonomous aircraft, named Aquila, 
which is capable of flying at an altitude of 18-20km over a defined trajectory to create a communication coverage region of about 100\,km. 
Aquila is self-powered through solar panels integrated on wings wider than a Boeing 737 and has light weight to increase the flying time. 
Moreover, it counts with a control system to adjust the GPS-based route and monitoring the most important flying parameters (like heading, altitude, airspeed, etc ...), 
and implements propellers able to operate at both low and high altitudes, 
thus at different associated air densities. Aquila employs free space optic (FSO) links to connect ground access points, 
which in turn serve ground users using either Wi-Fi or LTE technology. Facebook Aquila system architecture is illustrated in Fig.~\ref{fig:AquilaSketches}. 

\subsection{Google Loon}
\begin{figure}[t]
	\includegraphics[scale=0.35]{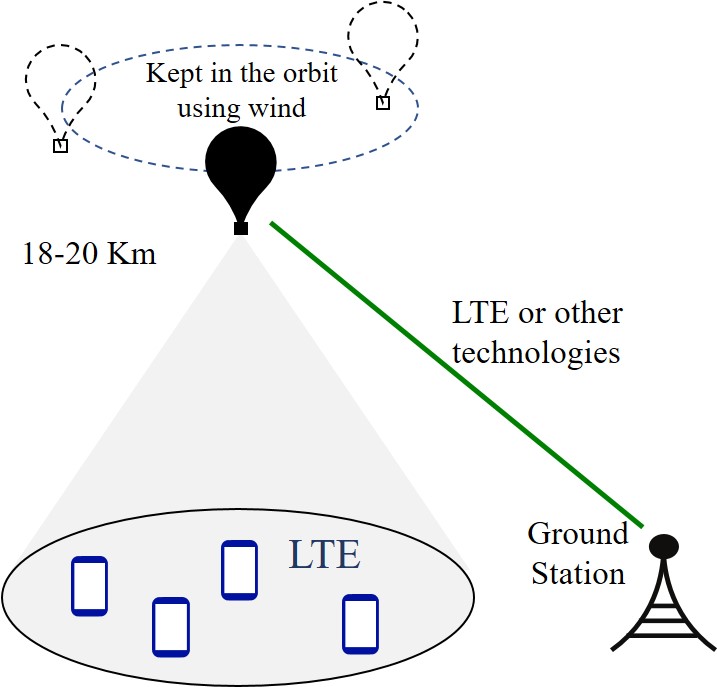}
	\centering
	\caption{Google Loon system architecture.}
	\label{fig:LoonSketches}
\end{figure}

Similar to Facebook Aquila, the Google Loon project~\cite{LoonProject} aims at bringing Internet connectivity in remote areas. 
This is achieved by adopting stratospheric balloons to relay radio communication links from ground stations to users' LTE phones out of the coverage of traditional ground cellular communication infrastructures. 
However, Google's Loon has a number of features that makes it different to Facebook's Aquila. 
The first one is that the communication may be directly relayed to the end user and not to ground access points. 
The second one is associated to the way the positioning of the balloons in the sky is controlled to generate the required coverage area at the ground. 
Instead of using propellers to maintain a pre-defined route,
they appropriately adjust their altitudes taking advantage of the stratified wind currents in the stratosphere. 
In fact, each layer of stratosphere is associated to a different wind direction and speed which can be monitored through machine learning algorithms and used to keep the balloon around the ideal location. 
Fig.~\ref{fig:LoonSketches} shows the system architecture implemented by the project. 

\subsection{Nokia F-Cell} \label{subs:nokiaprototype}

An interesting example of drone base stations prototyping for low altitude applications is the Nokia F-Cell project~\cite{nokiafcell}. 
The fundamental problem that F-Cell tries to solve is the high cost associated to the deployment and installation of a large number of small cells.
F-Cell is an innovative solar-powered, self-configured and auto-connected drone deployed small cell served by a massive MIMO wireless backhaul. 
The F-Cell architecture is comprised of a closed loop, 64-antenna massive MIMO array placed in a centralized location that is used to spatially multiplex up to eight energy autonomous F-Cells, 
each of which has been redesigned to require minimum processing power and mount a solar panel no larger than the cell itself. 
The key innovations proposed by F-Cell can be summarized in the following three aspects:



\begin{figure}[t]
	\centering
	\subfigure[Nokia F-Cell design, including the carbon fibre external cradle and the internal hardware.]{\includegraphics[width=\columnwidth]{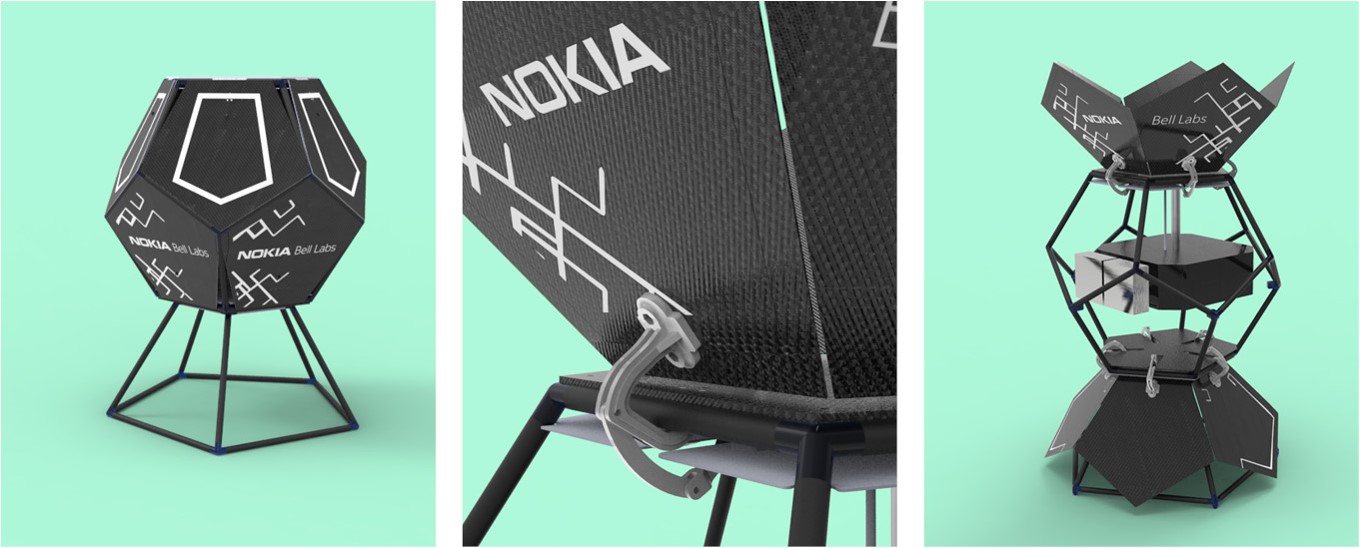}}
		\subfigure[Nokia F-Cell system architecture.]{\includegraphics[width=\columnwidth]{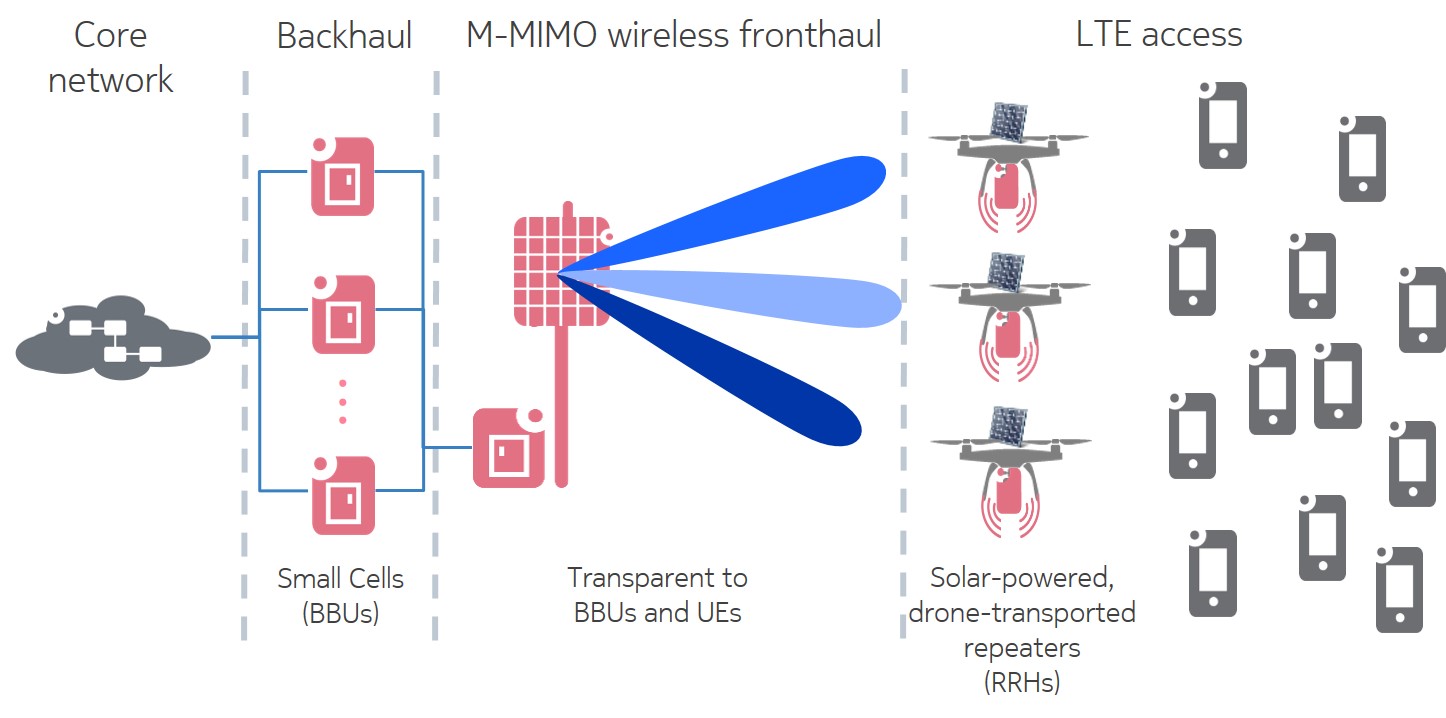}}
	\subfigure[Nokia F-Cell showcase performed at Sunnyvale, CA, USA.]{\includegraphics[width=\columnwidth]{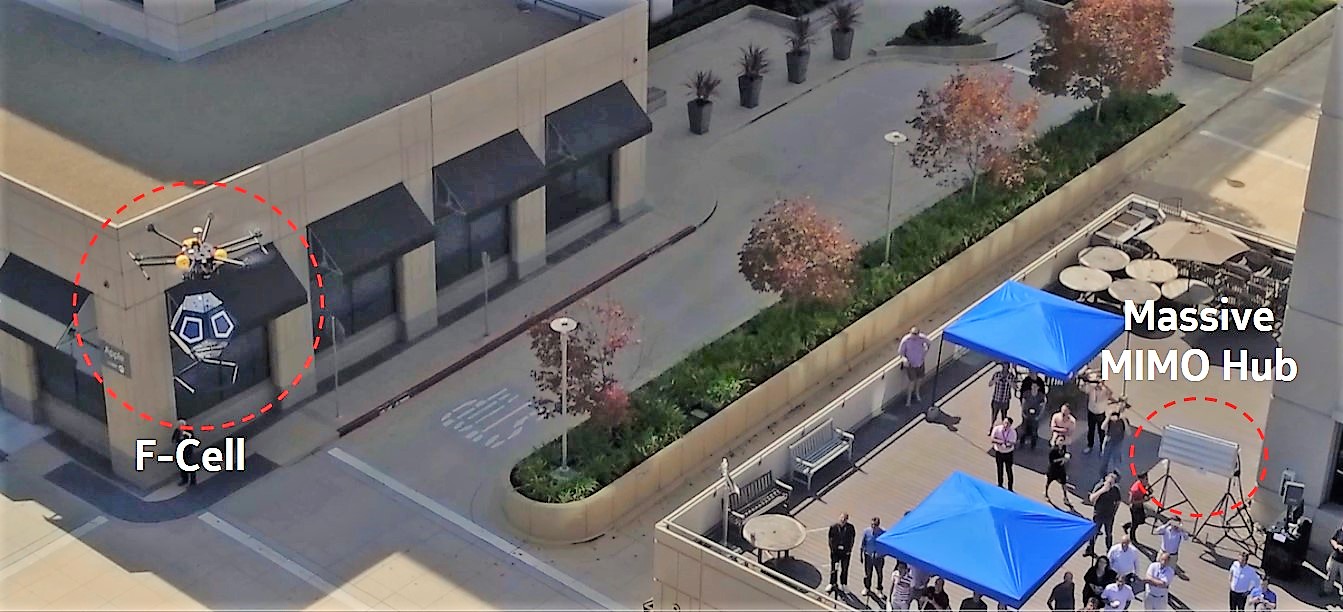}}
	\caption{Nokia F-Cell: (a) Design (b) System Architecture, and (c) Showcase.}
	\label{fig:NokiaFCellProj}
	\vspace{-0.3cm}
\end{figure}


\begin{enumerate}
\item \textit{Remove the need for a wired power supply} through the design of a energy-efficient small cell that reduces to the minimum the baseband signal processing, 
and the power consumption of the transceivers and their associated radio frequency hardware components to consume no more than 15 Watts. 
\item \textit{Remove the need for a wired backhaul} through the exploitation of massive MIMO spatial multiplexing capabilities below 6 GHz to provide high throughput wireless links in non-LOS conditions. 
\item \textit{Remove the constraint of a fixed deployment} through the flexibilities introduced at point 1) and 2), together with an optimized design of the small cell, which weighs less than 6 Kg. 
Fig.~\ref{fig:NokiaFCellProj}(a) illustrates the design of the F-Cell prototype, 
including the external carbon-fibre cradle with integrated solar panels and the internal box containing both hardware transceivers {\revise{and}} processing boards. 
Overall, this allows the transportation and relocation of small cells anywhere and at any time.
\end{enumerate}

In order to remove these constraints, 
F-Cell promulgates a novel system architecture, 
which is illustrated in Fig.~\ref{fig:NokiaFCellProj}(b). 
The essential feature introduced by F-Cell was the combination of both 
1) a fully digital massive MIMO hub with 64 active transceivers, 
and 2) analog repeaters. 
This transforms the wired backhaul into a multi-stage wireless {\revise{fronthaul}} with sparse remote radio heads (RRHs).  
Fig.~\ref{fig:NokiaFCellProj}(c) presents a flying test demonstration of the F-cell project,
where the drone-transported F-Cell and the massive MIMO hub are highlighted in red. 
The proposed solution is completely transparent both to the baseband units (BBUs) and to the user equipments (UEs). 
Importantly, F-Cell supports non-LOS wireless networking in frequency division duplex (FDD) mode, 
where downlink (DL) and uplink (UL) transmissions are performed at different frequencies. 
This generally entails the need of performing complex signal processing tasks at the analog repeater during the channel estimation procedure~\cite{7402270_MassiveMIMOMythsMarzetta, 1608543_FastTrasnferMarzetta, 4527202_EchoMIMO_Warme, 6777295_TrainingFDDMassiveMIMO_Love}. 
However, in order to simplify the repeater and reduce both its power consumption and associated weight, 
F-cell implements a joint uplink (UL) and downlink (DL) channel estimation procedure \emph{only} at the massive MIMO hub side~\cite{FCellPatent_BlumeMarzettaKlein, 8025616_WesemannMarzettaFDDTraining, Yao2018mMIMO_WCNC}. 
Additional details of the baseband hardware (HW) and software (SW) processing architecture and operations implemented at the massive MIMO hub are presented in~\cite{7996696_FlexHWSW_Galati}.






\begin{table*}[t!]
	\setlength\extrarowheight{4pt}
	\centering
	\caption{Specifications of testbeds and real experiments in the literature}
	\label{tbl:testbed}
	\begin{tabular}{|m{4cm}|m{5.2cm}|m{2.8cm}|m{1.6cm}|m{2cm}|}
		\hline
		\textbf{UAV Model} & \textbf{Objective} & \textbf{Equipments} & \textbf{Protocol} & \textbf{Controller}
		\\
		\hline
		AR.Drone 2 &Communication between multiple UAVs~\cite{7208426} &Raspberry Pi (Model B)&802.11 &  Software controller (NodeJs)
		\\
		\hline
		Autonomous helicopter& Accuracy of the navigation system~\cite{Tuna201427,6421446}& -- & -- &  Software controller (C)
		\\
		\hline
		Skywalker X8 & Investigating the impact of altitude, and motion on communication quality~\cite{7063641}& --& 802.11g/ 802.11b & Manual remote control
		\\
		\hline
		Senior Telemaster airplane kit& Evaluating load-carry-deliver protocol by one UAV~\cite{4225050}& --& 802.11a/ 802.11g/ 802.11b & ---
		\\
		\hline
		Dragonflyer X6& Aerial localization~\cite{6361401}& Lenovo W500 laptop, Nokia N900& XBee & Software controller
		\\
		\hline
		NexSTAR & Controlling the UAV path to improve link quality between two mobile nodes~\cite{6214700}&--& 802.11b/ 802.11g & Software controller
		\\
		\hline
		Fixed-wing and rotor based UAVs & Validating the impact of UAV as a relay~\cite{7389838} &--&WiFi& --
		\\
		\hline
		IRIS quadrotor & Deriving the energy consumption model~\cite{7101619} &--&--& --
		\\
		\hline
		Firefly (Ascending Technologies)&Enhancing video streaming~\cite{Kacianka2015}&AscTec's Ubuntu 12.04&IEEE 802.11& Manual controller
		\\
		\hline
		Fixed wing UAV, captive balloon,&Comparing the analytical and simulation results of different channel models ~\cite{5700244}&ARM-7 CPU with Linux OS&WLAN&Manual controller
		\\
		\hline
		Commercial UAV &Distinguishing between terrestrial and aerial UEs~\cite{8292313,8269067} & QualiPoc1 Android smart phone, &LTE 800MHz&  
		\\
		\hline
	\end{tabular}
\end{table*}

\subsection{Eurecom Perfume} \label{subs:eurecomprototype}

In the context of a European Research Council (ERC) 5-years funding scheme, 
Perfume project has studied and developed the concept of "autonomous aerial cellular relay robots", 
where UAVs act as a relay base stations capable to enhance connectivity and throughput performance for off-the-shelf commercial terminals~\cite{PerfumeProject}. 
The key target of the Perfume project{\revise{ is to design machine }}learning algorithms able to find and constantly update the optimal 3D position of flying wireless relays using fine-grained information of their LOS conditions together with other radio measurements~\cite{7996921_EurecomPerfumeOptPositioning}. 


%
%

\subsection{Huawei Digital Sky}

Huawei's Wireless X Lab activated in 2017 the Digital Sky Initiative to boost trials of specific use cases involving connected drones. 
With this purpose, the city of Shanghai has created an end-to-end ecosystem populated by key stakeholders such as mobile operators, 
cloud management firms, delivery companies, public transportation institutions, or even air quality monitoring agencies. 
Two authorized flying zones of 6km diameter and with a maximum height of 200m have been created, 
with distributed wireless charging zones at the ground. 
In this ecosystem, cellular networks are involved to ensure C\&C between drones and ground control stations. 
In particular, one interesting use case is a remotely operated passenger carrying drone (a taxi drone), 
which is controlled through live high quality video streaming transmitted over a 4.5G cellular network directly to the operation room~\cite{TaxiDrone}.  


\subsection{Other relevant testbeds}

There exist a number of smaller testbeds that verify practical problems like the reliability of UAV communication links, 
the UAV battery consumption, 
or the impact of weather conditions. 
These testbeds are summarized in Table~\ref{tbl:testbed}, and below we expand about some of the most relevant ones:

\begin{enumerate}
\item In~\cite{7208426}, a realization of a low-cost test bed based on AR.Drone 2 and Raspberry Pi is presented. The main idea is to validate the possibility of creating a flying ad-hoc network based on 802.11 standard able to establish communication links among UAVs. 

\item An autonomous helicopter is also used in~\cite{Tuna201427,6421446} to investigate navigation system, comparing a pre-planned trajectory with the actual flight path by the helicopter. A maximum error of 3 meters in one square kilometre area of field experiment is reported. The navigation system is tested for a single UAV.

\item An experimental test when a UAV acts as a relay to transmit information from an underwater vehicle to a ground base station is described in~\cite{7063641}. In this setup, a UAV moves in a circular way around an underwater vehicle with the speed of 18m/s. The effect of altitude, trajectory radius, waves, and weather conditions are analyzed in the communication between the UAV and the underwater vehicle.
\end{enumerate}

%% file: Regulation.tex
\section{Regulation}
\label{s:reg}

\begin{table*}[]
	\centering
    \caption{Summary of the most representative current UAV regulations}
    \label{tbl:regulsummary}
	\begin{tabular}{|m{1cm}<{\centering}|m{1.8cm}<{\centering}|m{1.8cm}<{\centering}|m{5cm}<{\centering}|m{1.8cm}<{\centering}|m{1.8cm}<{\centering}|m{1.8cm}<{\centering}|}
		\hline
		\textbf{Country} & \textbf{Applicability} & \textbf{Technical Requirements} & \textbf{Operational Limitations} & \textbf{Administrative Procedures} & \textbf{Human Resources} & \textbf{Ethical Constraints} \\ \hline
		\textbf{Australia} \cite{casaadvisory} & classification: weight/purpose & N/A & \begin{itemize}
		    \item minimum distance to people: 30m \item height limit: 120m \item minimum distance to airport: 5.5km \item daytime only (not after sunset) \item by visual line of sight (VLOS) only \item cannot operate over popular areas \end{itemize}
		& insurance strongly recommended & $>$2kg: pilot's license required & respect personal privacy \\ \hline
		{\revise{\textbf{Canada} \cite{canadauav}}} & {\revise{classification: weight/purpose}} & {\revise{N/A}} & {\revise{\begin{itemize} \item daytime only and not in clouds \item marked with pilot's name, address and telephone number\item maximum distance to pilot: 500m \item height limit: 90m and VLOS only \item minimum distance to vehicles, vessels and the public: \begin{itemize} \item 30m (250g$<$UAV's weight$\leqslant$1kg) \item 76m (1kg$<$UAV's weight$\leqslant$35kg) \end{itemize} \item minimum distance to heliports/aerodromes/natural hazard or disaster area: 1.9km/5.6km/9km \item away from controlled or restricted airspace and do not interfere with police or first responders\end{itemize}}} & {\revise{Special Flight Operations Certificate required when flying drones for work or research, or the drone weighs over 35kg}} & {\revise{N/A}} &{ \revise{respect the privacy of others}} \\ \hline
		\textbf{Chile} \cite{chiledac} & classification: weight & emergency parachute required & \begin{itemize} \item minimum distance to people: 20m (vertical); 30m (horizontal) \item height limit: 130m \item maximum take-off weight: 9kg \item by VLOS and daytime only \end{itemize} & flight authorization required & remote pilot's license required & respect the privacy of others\\ \hline
		\textbf{China} \cite{caac} & maximum weight: 7kg & N/A & \begin{itemize}
		\item minimum horizontal distance to other aircraft: 10km \item minimum vertical distance to other aircraft: 600m (altitude$\leq$8400m); 1200m (altitude$>$8400m) \item maximum speed: 120km/h \item daytime only and by VLOS only \end{itemize} & operational safety evaluation required & pilot certification required & N/A \\ \hline
		{\revise{\textbf{Japan}} \cite{japanuavsr}} & {\revise{classification: weight/purpose}} & {\revise{N/A}} & {\revise{\begin{itemize} \item daytime only \item by VLOS only \item minimum distance to people, other UAVs, ground properties and water surface: 30m \item cannot fly over event sites \item cannot carry hazardous materials \item cannot drop any objects \item prohibited airspace: \begin{itemize} \item 150m above ground level \item airspace around airports \item densely inhabited districts (DID) \end{itemize} \end{itemize}}} & {\revise{permission required for operation in the prohibited airspace}} & {\revise{N/A}} & {\revise{N/A}} \\ \hline
		\textbf{South Africa} \cite{sacaa} & maximum weight: 7kg; classification: purpose & N/A & \begin{itemize} \item height limit: 46m
		\item minimum distance to people and property (unless permitted): 50m \item minimum distance to airport: 10km \item by VLOS only \item in daylight and clear weather conditions \end{itemize} & air services license and operational certificate required & remote pilot's license required & respect privacy-by-laws\\ \hline
		\textbf{UK} \cite{caauk} & classification: weight/purpose & beyond visual line of sight (BVLOS): collision avoidance required & \begin{itemize}
		\item minimum distance to people: 50m \item height limit: 122m \item minimum distance to congested area: 150m \item by VLOS only (up to 500m) \end{itemize} & approvals vary for different operations & pilot competency required & protect data integrity and confidentiality \\ \hline
		\textbf{US} \cite{usfaa} & classification: weight/purpose & operation during civil twilight: anti-collision lights required & \begin{itemize} \item height limit: 122m, VLOS only
		\item minimum vertical distance to clouds: 152m \item minimum horizontal distance to clouds: 610m \item minimum distance to airport: 8km \item maximum speed: 161km/h \end{itemize} & aircraft registration required; $>$25kg: operational certificate required & remote pilot certification required & respect privacy-related laws\\ \hline
	\end{tabular}
\end{table*}

The evolution of UAV regulations should keep pace with the rapid emergence of UAVs, 
which significantly contributes to the integration of UAV into national and international aviation systems. 
In this section, socio-technical concerns of drone operations are outlined first. Then we explain the main criteria that constitute the current UAV regulation frameworks. 
Finally, the past and current status of UAV regulations are reviewed.

\subsection{Socio-Technical Concerns of Drones}
Emerging technologies have enabled the widespread use of drones and their strong operational capabilities. 
As a result, there are increasing concerns regarding privacy, data protection and public safety from national and international aviation authorities. 
To understand the motivation of the development of UAV regulations, socio-technical concerns of drones need to be analyzed.
\begin{itemize}
	\item \emph{Privacy:} The operation of drones can be a serious threat to the privacy of both individuals and businesses. 
For example, in the case that drones are employed for deliberate surveillance, 
they could intentionally violate individuals' and businesses' privacy. For missions such as aerial photography and traffic monitoring, privacy breaches can instead be unintentional~\cite{Liu2018PrivPic_GC}. 
Moreover, high maneuverability and sensitive on-board instruments have made drones even more capable of privacy breaches. 
For example, small drones with a low noise level can easily enter a private property without being noticed.\footnote{The DJI Mavic Pro Platinum has achieved 4 dB (60\%) noise reduction compared to the DJI Mavic Pro~\cite{Mavic}.} 
Indeed, images and videos taken by high definition (HD) camera could be streamed live~\cite{Liu2018PrivPic_GC}. 
Although every country has legislation to protect the privacy of the public's and citizens', 
such as the Commonwealth \textit{Privacy Act 1988} in Australia and the US Privacy Act of 1974~\cite{sibthorpe1995record}\cite{langheinrich2001privacy}, 
these rules might be out-of-date due to the rapid development of emerging technologies. 
Therefore, the operation of drones needs to be regulated to further protect the privacy. 
	\item \emph{Data Protection:} During their operation, drones are usually equipped with sensors that collect personal data such as images, videos and location data. 
How these personal data will be processed, used, stored and disclosed should concern public institutions. 
According to data protection laws, citizens' personal information should be protected from abuses~\cite{finn2014study}.
Invisible and indiscriminate data collection capabilities of drones contribute to infringe data protection rules. 
``Invisible'' refers to the fact that drones can secretly collect data due to their aerial capabilities and sensitive equipment on-board such as high resolution and night vision camera. 
Then collected data could be immediately uploaded online or transferred to a location that is distant from the data subject. 
Therefore, it is difficult for data subjects to be aware of the leakage of their personal information. 
Moreover, as a result of high mobility, drones indiscriminately collect and store a mass of data, 
which is against data protection principles~\cite{finn2014study}. Consequently, operation of drones should also be governed to protect personal information.  
	\item \emph{Public Safety:} Public safety is another major concern for drone operations. Compared to traditional manned aircraft, drones are usually insufficiently maintained and more likely to encounter pilot errors. 
As a result, drone operations are faced with higher safety risks. 
According to~\cite{bolkcom2003unmanned}, 
the accident rates for UAVs are significantly higher than those of manned aircraft. UAV accidents include collisions with manned aircraft or terrain:
	\begin{itemize}
\item \emph{UAV collisions with manned aircraft} might lead to engine shutdowns or damaged surfaces, 
risking the loss of control. 
Therefore, in many countries there are constraints such as maximum allowed flight heights and minimum distances to airports for drones' operation~\cite{canis2015unmanned}. 
\item \emph{UAV collisions with terrain} usually cause loss of control, 
which might hurt civilians on the ground. 
Hence some countries forbid drones to fly over certain areas such as specific urban areas with high population density~\cite{canis2015unmanned}.
 \end{itemize}	
\end{itemize}

\subsection{Criteria}
\label{sec:regulationCriteria}
Based on the socio-technical concerns specified above, UAV regulations are framed and developed. Current UAV regulations are mainly based on six criteria, as explained below~\cite{stocker2017review}. 
\begin{itemize}
	\item \emph{Applicability:} Applicability describes the scope that UAV regulations apply to. For example, drones are usually classified into groups based on weight or purpose, which might be treated differently by UAV regulations. 
	\item \emph{Technical requirements:} Technical requirements specify the mandatory instruments or techniques for drones. For example, collision avoidance mechanism could be a typical one.  
	\item \emph{Operational limitations:} UAV's operation is usually restricted by many factors. Typical operational limitations include maximum flight height, minimum distance to airport and individuals, prohibited areas, etc.
	\item \emph{Administrative procedures:} Certain procedures and documents might be required before a UAV is allowed to operate, which include registration, operational certificate and insurance. 
	\item \emph{Human resource requirements:} For certain categories of UAV and purposes of operation, the pilot needs to be qualified. 
	\item \emph{Implementation of ethical constraints:} This criterion appoints the demands for data and privacy protection when operating drones. 
\end{itemize}

\subsection{\revise{UAV Communications Regulation}}

{\revise{The potential of UAV applications and their implications in terms of reliability requirements and coexistence with other systems have not gone unnoticed by communications-related regulatory bodies. Below we summarize the most relevant advances in this regard:}

\begin{itemize}
	\item {\emph{The Electronic Communications Committee (ECC) within the European Conference of Postal and Telecommunications Administrations (CEPT)} formed a correspondence group on spectrum requirements for drones in Dec. 2015. This group produced the ECC 268 report discussing the technical, 
regulatory aspects and the needs for spectrum regulation for UAVs in Feb. 2018~\cite{ECC_268}. 
This report paves the way to the harmonisation across Europe of the frequencies dedicated to UAV communications. 
The use of cellular networks for UAV C\&C communications is addressed in this report and is currently under further study. 
Additionally, ECC 268 discusses the communications requirements of both professional UAV use cases, 
which could benefit from using individual licensed spectrum, 
and non-professional UAV applications, 
where unlicensed bands may suffice for short range communications.}
	
	\item {\emph{The Federal Communications Commission (FCC)} in the U.S. received in Feb. 2018 a petition for rule-making from the Aerospace Industries Association (AIA) ~\cite{FCC_Petition}. 
The petition is seeking to allow the secure communication of C\&C and non-payload data between UAVs and licensed pilots in the 5030-5091 MHz band. The FCC has to gather public comment before adopting a decision.}
	
\end{itemize}}

\subsection{UAV Flying Regulation: Past and Present}

\begin{itemize}
	\item Past: The first UAV regulation was proposed in 1944, 
right after the World War \uppercase\expandafter{\romannumeral2}. 
The first internationally recognized aviation regulation, 
the Chicago Convention, 
pointed out that the operation of UAVs should be authorized to ensure the safety of manned civil aircraft~\cite{sheehan1950air}. 
Since 2000, due to the rapid development of UAV and its increasing popularity, UAV regulations have evolved both nationally and internationally. 
In 2002, the United Kingdom and Australia first published their UAV regulations. 
In 2006, the International Civil Aviation Organization (ICAO) announced that it was necessary to issue an internationally acknowledged legislation for civil operations of UAVs. 
Since 2012, an increasing number of countries have established their own UAV regulations. 
	\item Present: Table~\ref{tbl:regulsummary} examines the current UAV regulation frameworks of {\revise{eight}} countries based on the criteria detailed in Sec.~\ref{sec:regulationCriteria}. 
Notice that UAV operations are currently prohibited in some countries such as Egypt and Cuba~\cite{bcfd}. 
{\revise{As it could be seen from Table~\ref{tbl:regulsummary}, every country has specific operational limitations to preserve the safety of both the public and UAVs. 
Moreover, most countries require the operation of UAVs to respect individuals' privacy. 
However, currently only the UK is aware of protecting the data collected by UAV operations. 
It is worthy mentioning that some countries are also proposing new drone regulations to address the safety demand, 
increasing popularity and economic significance of drones, such as Canada~\cite{canadauav}.}} 
Besides international and national aviation authorities, large enterprise groups are also making effort to help develop safe operations for drones. 
Recently, waivers for regulations have been given to Apple, 
Microsoft and Uber for their drone-testing projects, 
which will help the Federal Aviation Authority (FAA) shape the future development of UAV regulations in the US~\cite{amuamz}.  

\end{itemize}


%% file: Security.tex
\section{Security}
\label{s:security}

Security is a very important issue for any digital system. 
For a UAV-aided wireless communication system, due to its unmanned nature and required remote wireless communication, 
security is an even more serious problem. 
For example, compared to terrestrial BSs, 
if a flying cellular BS is compromised by attackers, 
then its serving UEs are more likely to lose cellular connections since the UAV may directly crash. 
Moreover, cellular UEs served by terrestrial BSs might suffer from strong interference due to LOS links,
if a UAV is manipulated by attackers. 
Therefore, it is significant to ensure the security of UAV systems when drones are used for cellular communications. 
In this section, cyber-physical security of UAV-assisted cellular communications is discussed.

\subsection{Cyber Security}
Since 2007, an increasing number of cyber-attacks to the UAV systems have been reported due to popularity of drones~\cite{javaid2012cyber}. 
When launching cyber-attacks, 
adversaries target the radio links of the UAV systems, 
which carry information such as data requested by cellular UEs, 
control signals and global positioning system (GPS) signals for UAVs' navigation. 
For example, with interception of these information, 
adversaries are able to steal data transmitted and requested by drones or even directly manipulate operating drones by taking advantage of their control signals. 
Since both data and control signals are transmitted through the radio links, 
ensuring the security of these wireless communication channels has become an important aspect of the whole UAV system's security. 
In this subsection, we will first present the scenarios where drones are used for cellular communications. 
Then, we evaluate the overall risk levels of different UAV connection types, namely, satellite, cellular and Wi-Fi links. 
After that, we analyze and list potential attack paths and corresponding defense strategies.

\begin{figure}
    \centering
    \subfigure[UAV UEs.]{
    \begin{minipage}[b]{0.5\textwidth}\includegraphics[width=0.98\columnwidth]{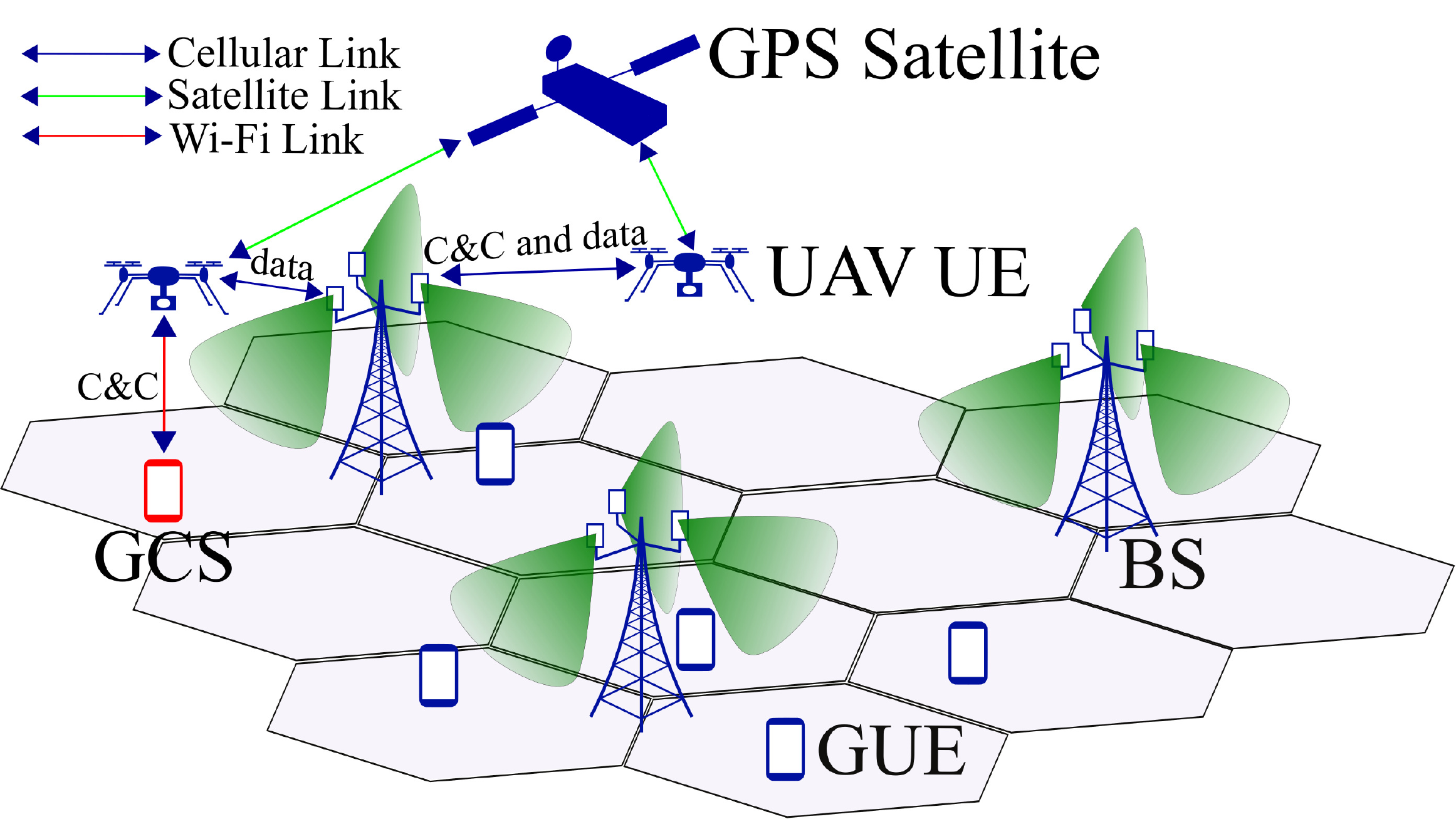}
    \end{minipage}
    }
    \subfigure[UAV BSs/relays controlled by terrestrial BSs.]{
    \begin{minipage}[b]{0.5\textwidth}\includegraphics[width=0.98\columnwidth]{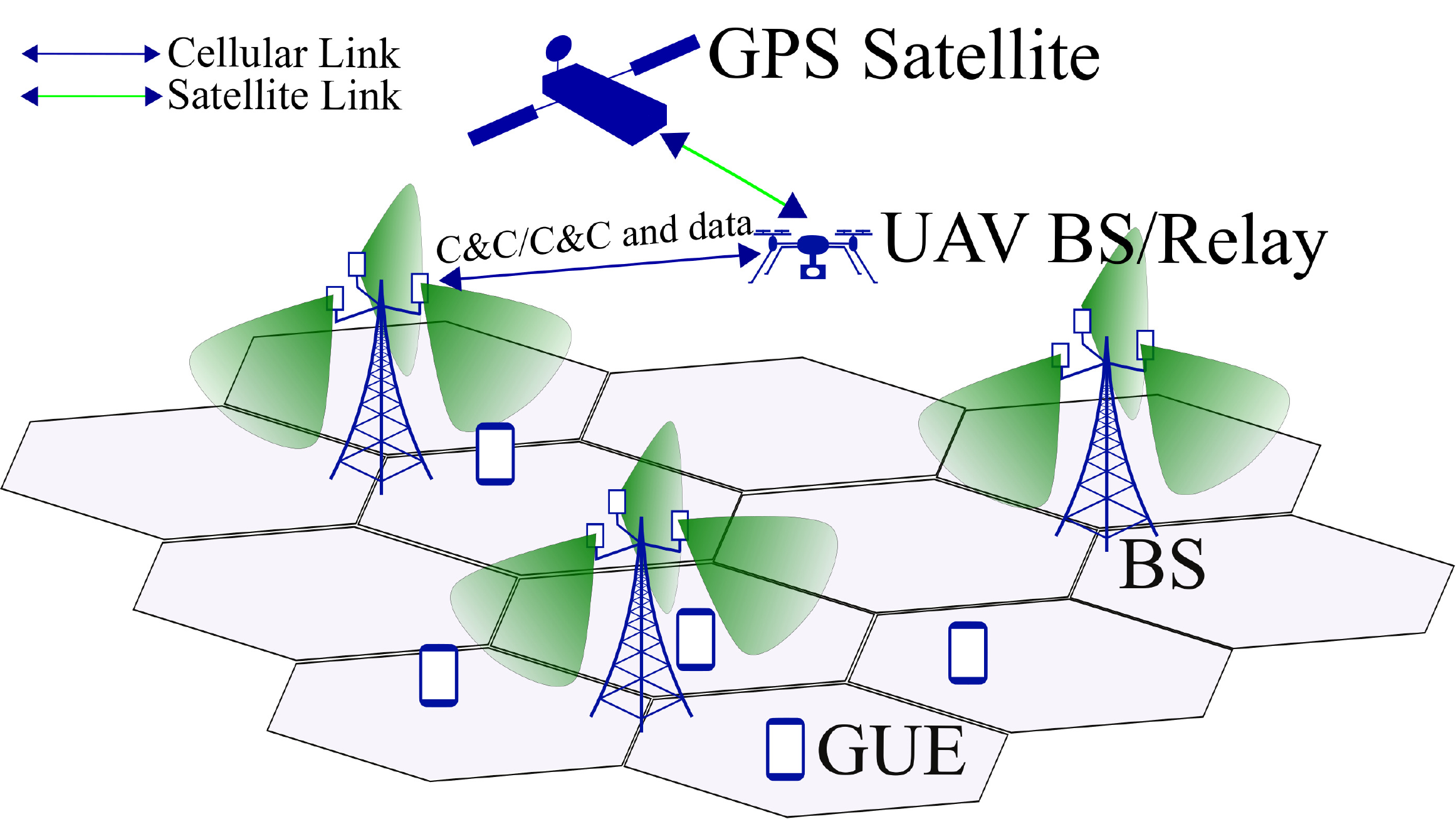}
    \end{minipage}
    }
    \subfigure[UAV BSs/relays controlled by third party GCSs.]{
    \begin{minipage}[b]{0.5\textwidth}\includegraphics[width=0.98\columnwidth]{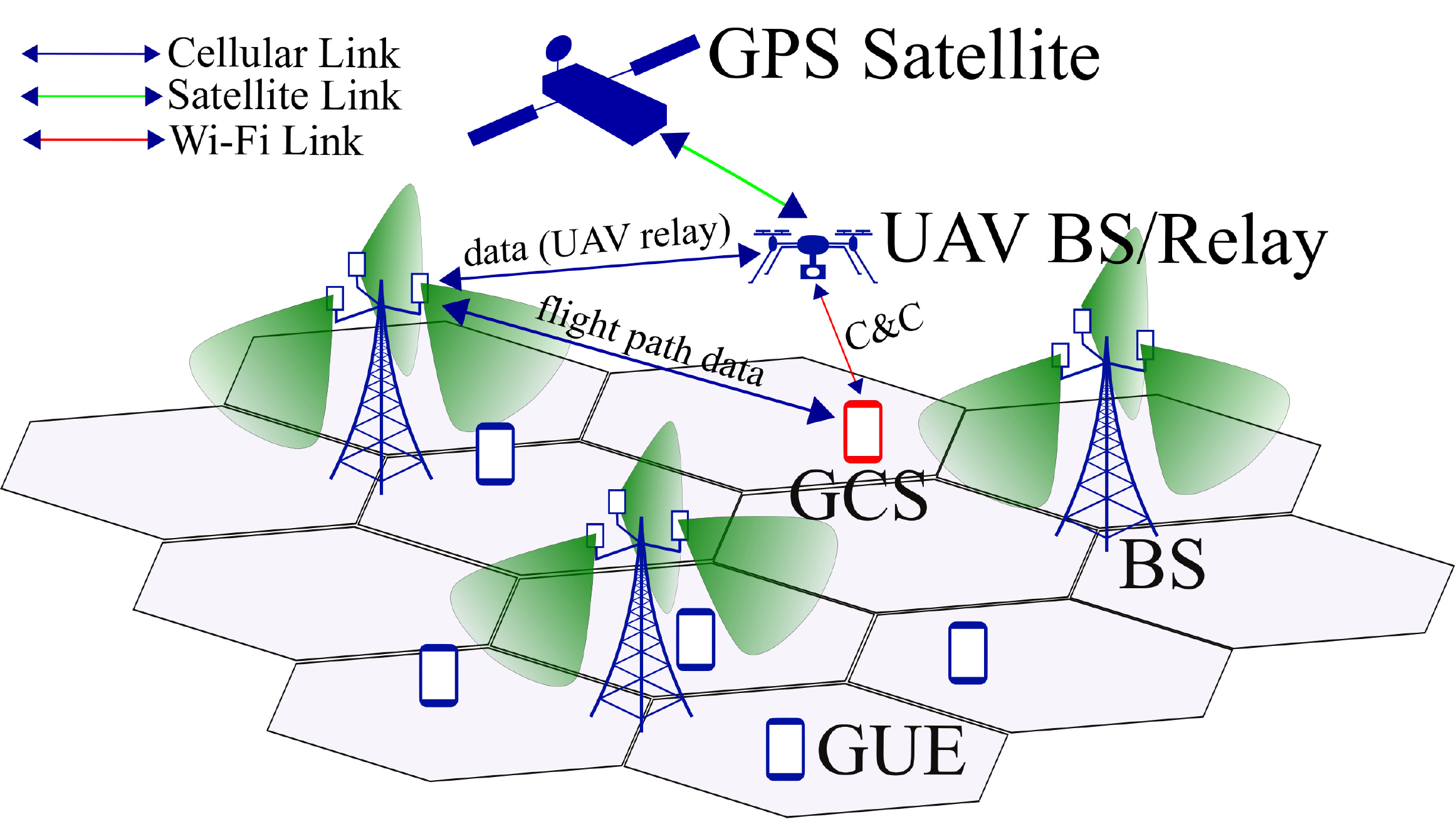}
    \end{minipage}
    }
 \caption{Use cases for cellular UAVs.} 
 \label{fig:UAV_use_case}
\end{figure}

\paragraph{Use Cases}
When drones are employed for cellular communications, 
they can serve as cellular UEs or flying cellular BSs/relays, as shown in Fig.~\ref{fig:UAV_use_case}. 
In case of UAV UEs, drones can be directly controlled either by terrestrial BSs or by ground control stations (GCSs) through non-cellular connections, 
which are mainly Wi-Fi connections~\cite{jawhar2017communication}. In the first case, both data and C\&C signals are transmitted via cellular connections. 
In the second case, data and C\&C signals use two separate radio links. 
Moreover, some drones are remotely controlled by GCSs via satellite connections,
such as the well-known Predator~\cite{ouma2011facets}. 
When drones serve as flying BSs/relays, 
similar conditions are present with respect to the previous UAV UEs case. 
However, in the presence of a third party GCS, flight path related data needs to be communicated by the cellular network to the GCS. Navigation information such as position, 
timing and velocity can be acquired from GPS satellites through satellite connections. 
As could be seen from above, 
there are three categories of radio links in these cases, 
which are satellite connection, 
cellular connection and Wi-Fi connection respectively.
It is known that compared to cellular networks and GPS networks, 
Wi-Fi networks are more insecure due to the unreliable and vulnerable security techniques {\revise{\cite{zeng2016wireless}}}. 
{\revise{For example, authors in~\cite{hooper2016securing} inferred that commercial Wi-Fi based drones are vulnerable to basic security attacks, which are even capable by beginner hackers. 
They demonstrated that by exploiting the standard ARDiscovery Connection process and the Wi-Fi access point, 
a flying Parrot Bebop 2 (its latest version) drone can directly crash~\cite{hooper2016securing}.}} 
Since GPS signals are broadcast and the signal format is specified to the public, 
it is easier to attack satellite connections than cellular connections where encryption keys and scrambling code are exchanged end-to-end.  
Therefore, 
risk levels of cellular connections, 
satellite connections and Wi-Fi connections are evaluated as low, medium and high, respectively. 
Table~\ref{table:Radio Link} summarizes the radio links and associated risk levels in the three cases above.

\begin{table}[t]
    \centering
    \caption{Risk analysis of radio links}
    \label{table:Radio Link}
    \renewcommand{\arraystretch}{1.5}
    \begin{tabular}{|c|c|}\hline
    \textbf{Role of Drones} & \textbf{Radio Link (Risk Level)}\\\hline
    \multirow{3}{*}{Cellular UEs} & Terrestrial BSs to drones: cellular (Low)\\\cline{2-2} &  Drones to GPS satellites: satellite (Medium)\\\cline{2-2} & Drones to GCSs: Wi-Fi (High)\\\hline
    \multirow{3}{*}{\tabincell{c}{Flying BSs/Relays\\ (Terrestrial BSs\\ controlled)}} &
    Terrestrial BSs to drones: cellular (Low)\\\cline{2-2} & \tabincell{c}{Drones to GPS satellites:\\ satellite (Medium)}\\\hline
    \multirow{5}{*}{\tabincell{c}{Flying BSs/Relays\\(Third Party GCSs\\controlled)}} & Terrestrial BSs to drones: cellular (Low)\\\cline{2-2} & \tabincell{c}{Terrestrial BSs to third \\party GCSs: cellular (low)}\\\cline{2-2} & Drones to GPS satellites: satellite (Medium)\\\cline{2-2} & Third party GCSs to drones: Wi-Fi (High)\\\hline
    \end{tabular}
\end{table}

\paragraph{Threat Identification and Countermeasure}
After assessing the overall risk levels of the radio links in the use cases above, 
threats for these radio links are identified and listed. Corresponding countermeasures are also presented. 
Finally, likelihood and impact of identified threats are evaluated in Table~\ref{table: Threat Analysis}. 
\begin{itemize}
	\item Jamming: Adversaries generate interference signals in the same frequency band to disrupt the reception process, 
which is a common way of integrity attacks. For example, GPS jamming has become a critical threat for drones. 
It was reported that in 2012 a small drone crashed and led to casualties, which was suspected to be caused by GPS jamming for the legitimate receiver~\cite{krishna2017review}. 
For jamming attacks, increasing the signal to noise ratio (SNR) could be a typical defense solution. 
However, this is always limited by how much power the transmitter can provide and how to lower the noise at the receiver by effective receiver algorithms. 
	\item Eavesdropping: Since both cellular and Wi-Fi connections employ wireless channels, 
adversaries might be able to obtain the transmitted information directly from the open environment. 
Eavesdropping breaches the confidentiality aspect of security. 
Encryption and physical layer security techniques could be used as a protective mechanism~\cite{Zhang2017phySec_TVT}. 
	\item Hijacking: Hijacking here refers to attacks that adversaries take over a radio link. 
For example, radio links between drones and GCSs in our scenarios are all Wi-Fi connections. 
To launch hijacking attacks, adversaries could first use {\it{deauthentication}} management frames to disconnect the association between a drone and the corresponding GCS. 
Then the drone can be remotely controlled by adversaries via 802.11 protocols. 
There are several security techniques against deauthentication attacks. 
Effective detection algorithms could be applied and transmitted frames could be encrypted. 
For example, WPA2 (802.11i-2004) encryption mechanism with proper key length is recommended as a countermeasure~\cite{he2017communication}. 
Moreover, dynamic secret key generation could provide a even stronger protection~\cite{Javali2015keyGen}. 
Alternatively, a Wi-Fi access point (AP) could be hidden by disabling service set identifier (SSID) broadcasting and Wi-Fi UEs could be restricted to those with certain media access control (MAC) addresses.  
	\item Spoofing: Adversaries can pretend to be some entity using false information. 
A typical spoofing attack to drones is GPS spoofing. By transmitting false GPS signals with higher power than the authentic ones, drones could be taken over by adversaries. 
To prevent GPS spoofing attacks, defense solutions such as jamming-to-noise sense and multi-antenna defense could be employed~\cite{borio2015real}\cite{magiera2015detection}. 
{\revise{Another example of spoofing attack could be address resolution protocol (ARP) cache poisoning attack. 
In~\cite{hooper2016securing}, authors successfully disconnected a drone from its controller by continuously sending spoofed ARP replies with the valid controller's MAC address.}} 
	{\revise{\item Denial of Service (DoS): For a DoS attack, adversaries will send excessive requests to the server, 
which causes network congestion. As a result, the legitimate users will lose their service. 
For example, in~\cite{hooper2016securing} authors employed the \textit{Telnet} application to fuzz the ARDiscovery (association between UAV and its legitimate controller) process} with simultaneous requests to become the controller for the UAV, which resulted in the crash of the UAV.} 
\end{itemize}

\subsection{Physical Security}
Besides cyber attacks, adversaries could also launch physical attacks to drones, which is another aspect of security concerns for UAV systems. 
To launch physical attacks, adversaries first need to obtain access to drones, which is achievable under two circumstances. 
First, adversaries can access a drone on the ground (damaged or ran out of battery) or capture a flying drone. 
Second, adversaries can control drones by successfully launching cyber attacks as introduced before. 
Here we summarize the attack paths and corresponding countermeasures according to attackers' capability levels~\cite{namuduri2017uav}. 
\begin{itemize}
	\item Low: Attackers aim to disassemble the captured drone to access its internal data, e.g. telemetry data via common interfaces such as USB. 
To defend such attacks, self-destruction mechanisms could be applied on drones, 
which will be enabled under pre-defined circumstances. 
However, self-destruction mechanism should only be triggered when necessary due to its strong side effects, 
e.g. potential threat to public safety, loss of both data and drones. 
	\item Medium: Attackers could access data through higher standard interfaces such as Joint Test Action Group (JTAG) and also access the embedded system. 
In this case, information stored on the drone needs to be encrypted. However, encryption may only delay the time taken by adversaries to obtain their desired data. 
	\item High: Adversaries are capable of launching advanced attacks such as side-channel attacks, 
fault injection attacks and software attacks to retrieve desired information from a drone. 
To deal with such attacks, superior cryptographic mechanisms and secure key management should be equipped by drones. 
\end{itemize}  

\begin{table}
	\caption{Threat analysis}
	\label{table: Threat Analysis}
	\centering
	\begin{tabular}{|c|c|c|}
		\hline 
		\textbf{Threat} & \textbf{Likelihood} & \textbf{Impact}\\
		\hline  
		Jamming & High & Low\\
		\hline
		Eavesdropping & High & Medium\\
		\hline
		Hijacking & Medium & High\\
		\hline
		Spoofing & Medium & High\\
		\hline
		{\revise{DoS}} & {\revise{High}} & {\revise{High}}\\
		\hline
	\end{tabular}
\end{table}

%% file: LessonLearned.tex
\graphicspath{{Figures/}}
\section{\revise{Lessons Learned}}\label{s:lessonlearned}

{\revise{
In this section, we summarize the key lessons learned during this survey, providing an overall picture on the current status of UAV cellular communications.
\begin{enumerate}[label= \emph{Sec. \Roman*:}, itemindent=2.5em]
	\item \emph{UAV types and characteristics.} The variety of UAV models and associated features is nowadays already very wide. From consumer low-cost drones with limited capabilities in terms of flying time and payload -- mainly suitable for recreational photo/video shooting applications and possibly involving a single wireless communication link towards a ground access point -- to commercial or military drones capable of travelling long distances and carrying heavy payloads -- mainly suitable for strategic operations like surveillance or wide area communication coverage involving high data throughput requirements and complex wireless communication apparatuses. However, although there already exists the possibility of finding a specific drone for almost every application, this does not imply that the pick represents an economically viable option. In our view, one of the main drawbacks of the current UAV technology is still the presence of a significant trade-off between flying time, carried payload and associated costs. In the future, a joint optimization of these three important metrics is desirable. Although the research community is already working in this direction, it seems that several years are still required to close this gap.
	\item \emph{Standardization.} All the major standardization bodies have already established dedicated study items and working groups to analyse the specific requirements for enabling reliable communications towards and from flying UAV through current cellular networks. 
Nevertheless, forecasting a future with a massive increase in the number of flying UAVs, classical network infrastructures, even with all the required extensions, might not be capable to contextually offer high quality of service to ground and aerial users. This is because they have been mainly designed to provide 2D coverage on the ground. With the introduction of aerial users, future cellular communication systems must embrace also the third dimension, with BSs able to point towards the sky and maybe dedicate specific resources to this type of new users. For this reason, massive MIMO 5G technology is foreseen as a strong candidate for introducing the required 3D spatial communication flexibility, thus pushing for an increasing focus in the 3GPP standardization activity to enhance and complement massive MIMO with UAV-dedicated solutions.
	\item \emph{Aerial base stations.} The widespread of UAV technology has generated a lot of interest in the possibility to deploy BSs where and when needed. In theory, we foresee benefits such as the avoidance of over-provisioned fixed network infrastructures to cope with hardly predictable data traffic peaks in time and the reduction of CAPEX and OPEX associated to site acquisition and maintenance or wiring. However, in reality, multiple issues need to be addressed before considering aerial BSs a cost-effective solution for replacing (even partially) conventional ground BSs. Current major concerns and research activities are focused on optimal placement and mobility of aerial BSs, power efficiency, recharging, and security.  
	\item \emph{Prototyping and field tests.} Several organizations have focused on verifying in realistic conditions the above-mentioned drawbacks and constraints associated with aerial BSs. Innovative approaches include those using futuristic autonomous solar-powered aircrafts, autonomous balloons, and self-powered aerial small cells. Although we find all these solutions extremely interesting and inventive, little can be said on their effectiveness in a wider cellular architecture with a mixture of aerial and ground BSs. Moreover, none of them have proven to be a cost-effective solution ready for wide adoption into an imminent product. 	
	\item \emph{Regulation.} Not only technical challenges are associated with UAV communications, but also the one related to privacy, public safety, administrative procedures, and licenses. Rules have been put in place in the majority of the countries in the world with the aim to control and limit the use of drones. Although different countries apply similar regulations, mainly concerning flying height, weight and safety distance from people, we are still far from an unified view on a basic common set of rules to be adopted world wide.
	\item \emph{Security.} With increasing number of drones flying in the sky, security becomes an extremely important requirement for drones not only to prevent actual falling and injuring people, but also to protect the data they are capturing and transferring to the ground network against possible hijacking. While it is clear that the problem can be significantly alleviated by adopting top notch software and hardware technologies, we should be mindful about the additional cost these might introduce to cellular-connected UAVs.  
\end{enumerate}
}}

%% file: FutureDirection.tex
\graphicspath{{Figures/}}
\section{Future Research Directions} 
\label{s:future}

In this section, we discuss the future research directions of UAV cellular networks. 
 
\subsection{UAV Simulator}

Real experiments with UAVs are inherently difficult due to tough regulations and need for large open space. 
Consequently, majority of researchers resort to simulations to evaluate the performance of their proposed systems and algorithms. 
These simulations often assume that UAVs can move at any direction at any time without any specific constraints of obstacles or any hardware restrictions. 
As a result, such simulation results may be far from realistic for many specific scenarios. 
Indeed, recent experiments with DJI quadcoptor drones have revealed that there are specific hardware limitations in terms of lateral acceleration, 
which prevents the drone from making turns at arbitrary angles~\cite{7974336}. 
It would be useful to develop publicly available simulators for UAVs, 
which would allow researchers to accurately simulate different types of drones according to their hardware specifications and subject to location-specific obstacles. 

There exists sophisticated open license simulators for ground vehicles, 
such as SUMO~\cite{SUMO}, 
which allows researchers to precisely simulate the microscopic movements of each vehicle on the road subject to road restrictions available from open maps, 
such as  OpenStreetMap~\cite{openstreet}. 
Currently, these maps only show the ground-based structures, such as roads and traffic lights. 
A future direction could be to add extensions to these maps which include details of other obstacles, such as high-rise buildings, roof-top cranes, etc.,  
which could affect UAV mobility in urban environment. 
Simulators such as SUMO could be extended to simulate microscopic mobility of UAVs of different makes and models. 
These extensions would allow UAV communications researchers to enjoy similar level of simulation support currently availed by researchers working on vehicular communications. 
It is worth noting that some drone vendors are offering their own simulators, 
such as  DroneKit from 3D Robotics~\cite{3dr}, 
Sphinx from Parrot~\cite{parrotsimul}, 
and DJI Assistant from DJI~\cite{djisimulator}, 
which allow researchers to connect a \textit{propelar-less} drone to the laptop and then collect drone telemetric data without actually having to go out and fly the drone. 
However, these simulators are still tied to the real drone, which still need to be purchased and connected to the laptop. 
As such, they do not provide the full benefit of a simulator, which can be used widely by anyone without having to purchase drones.



%
%


\subsection{Advanced UAV Mobility Control Based on Image Processing and Deep Learning}
Most consumer drones are equipped with high-fidelity cameras. 
With image processing and deep learning, UAVs can be programmed to identify the optimal hovering location or the optimal flying direction which would provide the best signal propagation between the UAV and the target ground BS (when UAV is acting as an aerial UE), 
or between the UAV and a ground UE (when the UAV is acting as an aerial BS), 
while considering real-life obstacles, 
such as buildings, rooftop cranes etc. 
Furthermore, energy status estimation and potential battery charging operations should be considered in advanced UAV mobility control as well.

\subsection{UAVs Antennas}
Because of the drones' ability to move in any direction with different speed, 
a new antenna design for airborne communication is required to achieve high data rate. 
One alternative to have high data rate transmission between UAVs and ground base stations is to have a tracking antenna installed on UAVs. 
The gyro, accelerometer and GPS information are utilized in order to track the ground station and tilt the antenna accordingly~\cite{7113485}. 
Moreover, 
limited space is another concern for installing antennas on UAVs~\cite{musselman2017antenna}, 
specially for small UAVs. 
A tilted beam circularly polarized antenna is proposed to install on the bottom of UAV~\cite{7821251} to save space. 
Simulation results showed that high performance in terms of return losses, 
axial ratio and radiation pattern can be achieved using such antenna.

\subsection{Aerial UE Identification}
One of the main challenges in introducing aerial UEs in LTE and 5G is identifying that an aerial UE has the proper certification to connect to the cellular networks. 
To this end, 3GPP proposed both UE-based and network-based solutions to indicate that a UE is airborne.
In the UE-based solutions, 
the UE can report informations such as altitude, flight mode and etc. 
With network-based techniques, different characteristics of the UE such as the mobility history, handover, etc. can help in UE detection. 
As stated before, since having a LOS is more probable for aerial UEs, 
they experience different radio conditions and interference than ground users. 
In this regards, 
three different machine learning techniques are employed in~\cite{8269067,8292313} to detect the presence of aerial UEs using standard LTE radio measurements, 
such as RSRP (Reference Signal Received Power) and RSSI (Received Signal Strength Indicator). 
The aerial UEs can be detected by up to 99\% accuracy through these methods. 
Developing more advance and intelligent algorithms, 
and utilizing various characteristics in future will lead in more precise aerial UE detection. 


{\revise{

\subsection{ Overcoming the Issue of Physical Reliability of UAVs}

By nature, UAV-BSs are expected to suffer from a physical reliability issue that does not concern existing terrestrial BSs. 
For example, UAVs can unexpectedly run out of battery, 
lose control due to sudden wind gusts, 
or even collide with other airborne objects. 
Such physical reliability issues will eventually affect the quality of wireless service and the overall quality of experience for the users. 
The physical reliability issue therefore must be addressed adequately before UAVs are integrated into cellular systems. 
It may take a long time before we can expect highly reliable UAVs systems at minimal cost. In the meantime, physical reliability must be factored in the optimization models when designing UAV systems. 
For example, the placement optimization for UAV relays or BSs  can include some physical failure and redundancy factor to improve overall network availability.  

\subsection{Mobile Edge Computing with UAV-BSs}

Cellular networks are moving towards a new paradigm called \textit{mobile edge computing} where the BSs provide not only communications, 
but also computing services to mobile users. 
This paradigm essentially brings the so-called cloud services closer to the mobile users thus reducing the latency for many real-time compute-intensive applications, such as augmented reality, speech recognition, and so on~\cite{edgecomp}.  

Several issues arise when edge computing is supported through UAV-BSs. 
To support computing services, 
UAV-BSs will have to be fitted with significant computing platforms, 
such graphical processing units (GPUs), which will increase both payload and energy consumption of UAV. 
Energy optimizations for UAVs therefore will have to factor in the computing tasks as well to ensure that UAV batteries can last longer. 
Computing session continuity is another issue that will have to be considered by the UAV path planning optimizations as the mobility of UAV-BSs can cause serious disruptions to ongoing computing tasks for mobile users. 
Indeed, researchers have realized such challenges and started exploring edge computing challenges and opportunities for UAVs~\cite{edge_jeong2018}.    

\subsection{UAV-supported Caching}

With the rapid development of Internet-of-Things (IoT)~\cite{Ding2018IoT_WCNC}, 
users' demand for multimedia data such as videos and images containing high data volume has significantly grown. 
Since the same popular contents are usually requested by many users at different times, 
wireless caching can relieve the pressure of data accessing by pre-caching the popular contents in local memory~\cite{wang2018power}. 
Owing to UAV's inherent high mobility, 
aerial caching provides a superior solution to deliver the contents more efficient than traditional static caching by addressing the user mobility problem. 
Authors in~\cite{chen2017caching} proposed the proactive deployment of UAV-supported caching to improve users' QoS in a cloud radio access network (CRAN). 
A very recent work investigated data popularity analysis based on machine learning, which can be employed in UAV-supported caching~\cite{Cheng2019caching_ToC}.
Authors in~\cite{8576651} also showed that UAV-supported caching could achieve higher multimedia data throughput in IoT system. 
Major challenges for aerial caching include location optimization of UAVs, user-UAV association, the contents to cache at UAVs and efficient power control strategy.  
}}

%% file: Conclusion.tex
\section{Conclusion}
\label{sec:conclusion}

We have surveyed complementary activities from academia, industry, and standardization on the important issue of integrating UAV into cellular systems. 
Our survey reveals that 3GPP has recently launched new study items to explore opportunities and challenges for serving UAVs using the existing 4G cellular networks. 
The outcomes of these preliminary studies point to some major interference issues arising from the height of the UAVs, 
but most of them can be addressed by deploying additional mechanisms in the existing 4G systems. 
It is expected that 5G and future systems will be better equipped to deal with UAV related challenges. 
We have found that a number of vendors have already built and demonstrated UAV-mounted flying base stations as a sign of their readiness to embrace UAV into the cellular systems. 
Our survey has identified a rapidly growing interest on this topic in the academic community,
which has resulted in a growing number of publications and workshops. 
There is an increasing activity within the regulation bodies to design and implement new regulations for UAVs to promote public safety and individual privacy.
Finally, 
we have identified new cyber-physical security threats, 
business and cost models, 
and future research directions for UAV-assisted cellular communication. 
We believe that UAV cellular communication is at a very early stage of development and we would expect to see continuing interest and progress in this exciting new research direction of cellular networking in the coming years.
